\documentclass[journal, 10pt, twocolumn, a4paper]{IEEEtran}

\usepackage{amssymb}
\usepackage{amsmath}
\usepackage{amsthm}
\usepackage{mathrsfs}
\usepackage{verbatim}

\ifCLASSINFOpdf

    \usepackage[pdftex]{graphicx}
    \graphicspath{{./}}
    \DeclareGraphicsExtensions{.pdf,.jpeg,.png}
	\usepackage[pdftex, bookmarks = true, colorlinks=false]{hyperref}
\else
	\usepackage[dvips]{graphicx}
	\graphicspath{{./}}
	\DeclareGraphicsExtensions{.eps}
	\usepackage[dvips, bookmarks=true, colorlinks=false]{hyperref}
\fi
\ifCLASSOPTIONcompsoc
	\usepackage[caption=false,font=normalsize,labelfont=sf,textfont=sf]{subfig}
\else
	\usepackage[caption=false,font=footnotesize]{subfig}
\fi

\RequirePackage{lineno}

\usepackage{color}
\usepackage[usenames]{xcolor}

\usepackage{arydshln,leftidx,mathtools}
\usepackage{booktabs}
\usepackage{subfig}
\usepackage{multirow}

\usepackage{array}
\newcolumntype{L}[1]{>{\raggedright\let\newline\\\arraybackslash\hspace{0pt}}m{#1}}
\newcolumntype{C}[1]{>{\centering\let\newline\\\arraybackslash\hspace{0pt}}m{#1}}
\newcolumntype{R}[1]{>{\raggedleft\let\newline\\\arraybackslash\hspace{0pt}}m{#1}}

\usepackage{paralist}
\usepackage{enumerate}

\DeclareMathOperator*{\diag}{diag}

\DeclareMathOperator*{\dilog}{Li_2}

\newtheorem*{theorem*}{Theorem}

\newtheorem*{definition*}{Definition}

\begin{document}
\setlength{\dashlinegap}{1.0pt}

\title{Detecting Human-induced Reflections using RSS of Narrowband Wireless Transceivers}
\author{
    \IEEEauthorblockN{H\"{u}seyin~Yi\u{g}itler, Riku~J\"{a}ntti and Ossi~Kaltiokallio}
    \IEEEauthorblockA{ \\\ Aalto University, Department of Communications and
Networking \\\ P.O. Box 13000, FIN-00076 Aalto, Finland
    \\\ e-mail:~\{name.surname\}@aalto.fi}
}
\maketitle

\begin{abstract}
Radio frequency sensor networks are becoming increasingly popular as an indoor localization and monitoring technology for gaining unobtrusive situational awareness of the surrounding environment. The localization effort in these networks is built upon the well-established fact that the received signal strength measurements vary due to a person's presence on the line-of-sight of a transmitter-receiver pair. To date, modeling this decrease in received signal strength and utilizing it for localization purposes have received a considerable amount of attention in the research field. However, when the person is in the close vicinity of the line-of-sight but not obstructing it, the signal reflected from the human body is also affecting the received signal strength and can be used for occupancy assessment purposes. In this paper, we first model the effect of human-induced reflections as a function of communication frequency, and then use the model as a basis for energy based occupancy detection. The derived methods are evaluated numerically and the detection probability of the proposed detector is validated with experimental data. The results suggest that when more than eight frequency channels are utilized, presence of a person can be detected using RSS measurements of a single transmit-receive pair with detection probability higher than 0.95 and false alarm probability less than 0.01 in an area of 2 x 2.5 m\textsuperscript{2}. Moreover, the important implications of the studied methods on the available narrowband radio frequency sensor network applications are discussed in detail.

\end{abstract}
\begin{IEEEkeywords}
Device-free localization, Low-cost wireless devices, signal detection, received signal strength measurements
\end{IEEEkeywords}

\section{Introduction}

Dense wireless sensor network deployments of low-cost and low-power nodes have recently been exploited for new sensing possibilities where the wireless propagation medium is probed. In general, these systems utilize received signal strength (RSS) measurements of the radio for gaining situational awareness of the surrounding environment and therefore, they are commonly referred to as \emph{RF sensor networks} \cite{Patwari2010}. These networks can be used for unobtrusive monitoring purposes such as device-free localization (DFL) \cite{Patwari2010} and non-invasive respiration rate monitoring \cite{Kaltiokallio2014}. An important advantage of RF sensor networks is that they can be realized using any wireless technology capable of measuring the RSS. The most common implementation platform is IEEE 802.15.4 based nodes \cite{Wilson2010, Kaltiokallio2012a}, however, WLAN \cite{Saeed2014} and RFID \cite{Geng2014} also enable RSS-based DFL. The technology has already been demonstrated for applications such as ambient assisted living \cite{Bocca2013}, residential monitoring \cite{Kaltiokallio2012a}, security and emergency surveillance \cite{Bjorkbom2013}, and roadside surveillance \cite{Anderson2014}.

RSS-based DFL is built upon the fact that people alter the propagation channel when they are close to the line-of-sight (LoS) of a TX-RX pair, resulting to different RSS readings at the RX \cite{Patwari2010, Wilson2010}. For such a localization system, there are four problems that need to be considered:
 \begin{inparaenum}[i.)]
	\item{\emph{modeling} the RSS measurements as a function of the person's position, geometry and electrical properties,}
	\item{\emph{detecting} the person's presence,}
	\item{\emph{classifying} how the person is altering the RF signals, and}
	\item{\emph{location inference} from the RSS measurements.}
\end{inparaenum}
The measurement models describe how people affect the propagation medium and relate this to the observed RSS. The detection problem aims to identify when a person is in the monitored area whereas classification seeks to identify the underlying cause of the measurement. The location of a person can be inferred from the RSS measurements by using an imaging approach referred as radio tomographic imaging (RTI) \cite{Wilson2010} or formulating the problem as a statistical inverse problem and solving it using sequential Monte Carlo methods \cite{Li2011}. In this paper, we aim at detecting a person in the monitored area using RSS measurements of a single TX-RX pair.
 
The RSS measurement models are closely related with the multipath propagation channel \cite{Patwari2010}. In a multipath rich environment, presence of objects in the propagation medium can alter existing or create additional multipath components while leaving some others unaffected. At the receiver, the observed variation on the received signal depends the propagation mechanism(s) affecting each component. The mechanisms altering them are: scattering, diffraction and reflection from the object's surface, transmission through the object's interior, and propagation losses \cite[ch. 3]{Vaughan2003}. For example, a person on the LoS (within the first Fresnel zone) causes the transmitted signal to experience a considerable loss known as shadowing \cite{Agrawal2009}. This observation can be explained by the combined effects of transmission through human tissue \cite{Kaltiokallio2014a} and diffraction \cite{Savazzi2014}. On the other hand, when the person is far away from LoS (outside of the first Fresnel zone), the impact of diffraction can be ignored \cite{Savazzi2014}, leaving scattering and reflection as the propagation mechanisms that need to be considered \cite{Patwari2011}. In this paper, it is assumed that the wavelength(s) of the measurement system is much smaller than the geometrical extent of the human body so that reflection can be considered as the sole dominating mechanism. Therefore, in this work we only consider human-induced reflections when the person is far away from LoS. The interested readers are invited to view the video clip accompanying our recent work \cite{Kaltiokallio2014a}, which shows the variation of measured RSS as a function of person's position \cite{ LinkLineVideo}. 

The RSS models also depend on whether the underlying measurement system is narrowband or wideband. These systems have fundamentally different models for the propagation channel \cite[ch. 7]{Molisch2011}, inevitably affecting the information available for detection and localization purposes. For wideband systems such as radar, and UWB and OFDM based communication systems; granularity of the measurements allows utilization of the well-known channel impulse response model. This model can be used for determining phase and amplitude of individual resolvable multipath components. On the contrary, the RSS measurements of narrowband systems is a function of phasor sum of all multipath components impinging on the RX antenna and therefore, individual components cannot be resolved. As a consequence, the RSS measurements of narrowband systems are not as informative for detection and localization purposes. In this paper, we consider narrowband systems, and thus, the detection methods developed for wideband systems cannot be utilized.

Detecting presence of a person in the monitored area is one of the outputs of DFL systems that can be utilized for occupancy assessment \cite{Naghiyev2014} in  applications such as people counting \cite{Lin2011}, residential monitoring \cite{Kaltiokallio2012a} or some home automation systems \cite{Mrazovac2012}. Typically, this is achieved by first localizing the person using human-induced RSS-variations due to shadowing, and then by making the occupancy decision. Since occupancy assessment only requires a binary input indicating a person's presence, detecting human-induced reflections independent of location inference enables different system optimization opportunities that are crucial for real-world DFL deployments.  
%
First, the computational complexity of the detector can be scaled for embedded implementations in the node to allow local decisions and distributed processing. 
Second, the power requirements of the DFL system can be decreased by duty-cycling when the environment is not occupied, which is a preliminary requirement of low-power deployments \cite{Yigitler2013}.
Third, the required number of nodes can be substantially reduced by detecting human-induced reflections 
since a TX-RX pair can monitor a larger area around the LoS. 
Therefore, detecting human-induced reflections is very important in RSS-based DFL and in this paper, the problem is further elaborated.

In this work, detection of human-induced reflections using RSS measurements of low-cost narrowband transceivers is studied. The effect of human-induced single bounce reflection on RSS measurements of coherent receivers is derived. It is shown that the effect is a periodic function of communication frequency with closed form trigonometric Fourier series coefficients. Based on this finding, an energy detector operating on RSS measurements at different frequency channels is introduced. Despite the fact that the benefits of utilizing multi-channel RSS measurements for localization purposes are known in DFL literature \cite{Kaltiokallio2012}, to the best of our knowledge, there is no published work modeling the impact of reflected signal as a function of frequency and utilizing the model for detection purposes. We further give a closed form expressions for probability of detection, false alarm, and the \emph{receiver operating characteristic} (ROC) of the introduced energy detector. The derived methods are evaluated numerically, and detection probability of the studied detector is validated with experimental data. The results suggests that when more than eight channels are utilized, the presence of a person can be detected using RSS measurements of a single TX-RX pair with detection probability higher than $0.95$ and false alarm probability less than $0.01$ in an area of $2 \times 2.5 \text{ m\textsuperscript{2}}$.

The remainder of the paper is organized as follows. In Section~\ref{sec:related_work} the related work is summarized. In Section~\ref{sec:background} the considered problem is defined and a mathematical model for human-induced reflections is introduced. In Section~\ref{sec:detection}, the energy detector which utilizes RSS measurements on the different frequency channels is presented and the numerical results are given. The probability of detection is validated experimentally in Section~\ref{sec:experimental_validation}. The implications and implementation possibilities are discussed in Section~\ref{sec:discussion} and thereafter, conclusions are drawn in Section~\ref{sec:conclusion}. Major notations used throughout the paper are given in Table~\ref{table:Major-Notations}.

\section{Related Work}\label{sec:related_work}

The detection of objects is the fundamental problem of radar \cite{Skolnik2008}, which usually utilizes a wideband measurement equipment for localization and detection \cite{Paolini2008}. Similarly, communication system such as UWB \cite{Kilic2014} and high-rate OFDM \cite{Zhou2013} provide fine grained information about the propagation medium that can be used for the same purposes. In this work, narrowband systems that cannot resolve individual multipath components are considered. 

Augmenting the RSS measurement models of narrowband systems to include reflections has been recently studied by Kaltiokallio \emph{et. al} \cite{Kaltiokallio2014a}. In the work, it is demonstrated that including reflections to the model enhances the localization accuracy of RSS-based DFL while size of the monitored area is increased. In this work, the detection problem is studied independent of location interference, and an energy based detector operating on RSS measurements at different frequency channels is introduced.

In RSS-based DFL, person's presence can be detected independent of location inference. The system proposed by Kosba \emph{et al.} \cite{Kosba2012} seeks for anomalies in RSS measurements to detect a person. The system refines the false alarm rate by cross validating decisions of different RSS sequences. The detection method proposed by Mrazovac \emph{et al.} \cite{Mrazovac2013} is based on candid covariance-free incremental principle component analysis. This method also uses a sequence of RSS measurements to first find the principle components, which are then used for calculating probabilities and information entropy to make the decision. Compared to the aforementioned works, the energy based detector studied in this work makes the following contributions. First, the detector has low computational overhead and memory footprint. Second, parameters of the detector are related to the reflected signal allowing one to attribute the decisions to a certain region of space. Third, the energy detector is solely based on RSS measurements on different frequency channels and it does not require a temporal sequence of RSS measurements. Thus, the algorithm can be implemented on constrained devices enabling fast decision about the presence of a person which in turn brings several advantages such as: decision refinement using different time samples, reduced communication protocol complexity, and enabling duty-cycling for low-power operation when the monitored area is vacant.

A person's effect on RSS is restricted by sensitivity region of the TX-RX pair. In terms of detectability, this region can be defined as the area where the person's presence can be detected with very high probability. In RSS-based DFL, the spatial models describing the sensitivity region are restricted to a very narrow region around the LoS \cite{ Wilson2010, Kaltiokallio2012a, Wilson2012, Zheng2012, Martin2014} and therefore, these models only account for shadowing. Such an approach necessarily implies that when the person is in the sensitivity region, the detection probability is $1$ within the entire region. Augmenting the RSS measurement models to include reflections renders new challenges including non-uniform and unconnected spatial probability distribution of detectability. Therefore, one of the most important results of this work is the spatial distribution of detection probability, which can be utilized for localization purposes and pre-deployment system optimization.

\begin{table}[t!]
\renewcommand{\arraystretch}{1.0}
\renewcommand{\tabcolsep}{1 mm}
\caption{Major Notations} \label{table:Major-Notations}
\centering
\begin{tabular}[c]{l p{6.8cm}}
\toprule [1pt]
$\mathbb{R}^m$	& $m$ dimensional Real and integer space. \\
$j$, $e$, $\hat{e}$	& $j=\sqrt{-1}$, and base of natural logarithm, and $\hat{e} = 10 \log_{10}(e)$. \\
$t$, $\tau$ & Time and delay in respective order. \\
$r(t)$, $s(t)$ & Received and transmitted signals. \\
$T_s$,$f_c$ & Symbol period and carrier frequency. \\
$c_0$, $\beta$ & Free space propagation speed and inverse wavelength, $\beta=f_c\big/c_0$. \\
$\alpha_i$, $\phi_i$, $\tau_i$ & Amplitude, phase and delay of $i^\text{th}$ multipath component of the received signal.\\

$\mathcal{P}_r$, $\mathcal{P}_s$& The power of received and transmitted signal. \\
$n(t)$ & Zero-mean complex Gaussian white noise with two-sided power spectral density ${N_0} \big/ {2}$.\\ 
$\Gamma$, $\eta$ & Reflection coefficient and path loss exponent. \\
$\epsilon_r$, $\theta$ & Relative permittivity and incidence angle. \\
$\boldsymbol{p}$, $\boldsymbol{p}_t$, $\boldsymbol{p}_r$ & A point in Euclidean space $\mathbb{R}^2$, $\boldsymbol{p} = [x~~y]^T$, 
														  and the positions of TX and RX in $\mathbb{R}^2$. \\
$d$, $\Delta$ & The distance between TX and RX, $d=\left\| \boldsymbol{p}_r - \boldsymbol{p}_t\right\|$,   
				and excess path length traveled by the wave reflected from point $\boldsymbol{p}$, 
				$\Delta = \left\| \boldsymbol{p} - \boldsymbol{p}_t\right\| + \left\| \boldsymbol{p} - \boldsymbol{p}_r\right\| - d$.\\
$\mathcal{P}_{LoS}$ & The LoS signal power. \\
$\mathscr{C}$, $C$, $\bar{\beta}$ & Channel set, number of channels, and average inverse wavelength in $\mathscr{C}$.\\
$\beta_l$, $\xi$ & Inverse wavelength of $l^{\text{th}}$ channel in $\mathscr{C}$, and $\xi = \beta_{l+1}-\beta_l$. \\
${z}(\cdot)$, $\zeta(\cdot)$, $\nu$ & Observation, the reflected signal function, and measurement noise in dB. \\
$\mathcal{P}_\zeta$, $\mathcal{E}_\zeta$ & Power and energy of $\zeta(\cdot)$. \\
$P_f$, $P_d$ & Probability of false alarm and detection. \\ 
$\mathcal{H}_0$, $\mathcal{H}_1$ & Null hypothesis and its alternative.
\\
\bottomrule [1pt]
\end{tabular}
\end{table}

\section{Background}\label{sec:background}

\subsection{System Overview}

\begin{figure}[ht]
\centering
 \setlength{\tabcolsep}{0pt}
\begin{tabular}{C{7cm}}
\subfloat[]{\includegraphics[width=6.6cm, height=2.8cm]{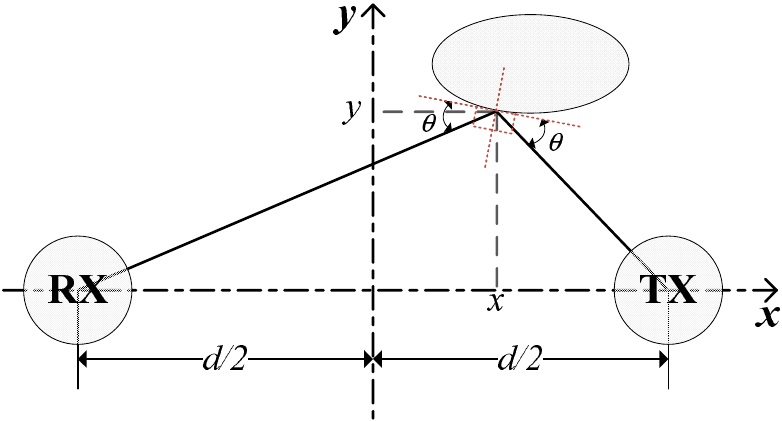}\label{fig:geometry}}\\
\subfloat[]{\includegraphics[width=7cm, height=2.6cm]{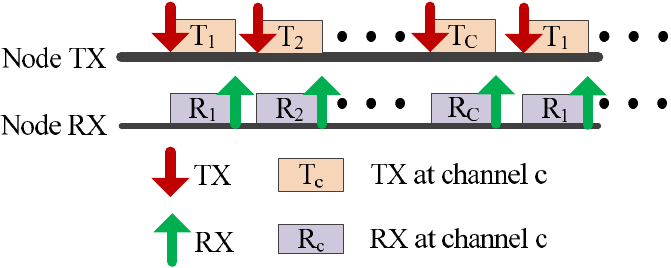}\label{fig:schedule}}  
\end{tabular}
\caption{In (a), overview of the studied scenario and the reference frame. In (b), the transmission schedule }
\label{fig:system_overview}
\end{figure}

In this work, the effect of human-induced reflections is studied. The emphasis of the paper is on detecting human-induced reflections using RSS measurements. In order to simplify the analysis, an idealized deployment scenario, illustrated in Fig.~\ref{fig:geometry}, is considered. The TX follows a multi-channel transmission schedule and the RX is programmed to listen the ongoing transmissions as shown in Fig.~\ref{fig:schedule}, which is the single transmitter version of the schedule discussed in \cite{Yigitler2013}.  

The narrowband communication system is presumed to fulfill the following assumptions. 
\begin{inparaenum}[i.)] 
\item{The wavelength is much smaller than average geometrical extent of the human body.}
\item{The nodes have coherent receivers so that the reference signal can be defined.}
\item{The RX always synchronizes to the LoS signal.}	
\item{The antennas are assumed to have an ideal frequency response and gain patterns, and they are vertically polarized.}
\item{The instantaneous in-phase and quadrature digital signals are not accessible but the RX is able to output one RSS measurement at each packet reception.}
\end{inparaenum}

In the analyzed scenario, a person is in the close vicinity of the LoS without obstructing it. Otherwise, the environment is assumed to be stationary so that the communication system only experiences slow fading. Under this condition, the transmission schedule shown in Fig.~\ref{fig:schedule} allows one to acquire RSS measurements altered by the same propagation channel if the time difference between measurements at the first and last frequency channels is small. For a single TX, it is easy to communicate over multiples of frequency channels and satisfy the aforementioned condition. On the other hand, for multiple transmitters the condition is satisfied by the schedule introduced in \cite{Yigitler2013} because each TX broadcasts sequentially on the different channels before changing the transmitting node. However, with large number of nodes and channels, the coherence time of the propagation channel is likely to be exceeded using the schedule in \cite{Bocca2013}. Therefore, if one is interested in similar analysis using multiples of transmitters and frequency channels, the utilized schedule should be selected appropriately.    

\subsection{Physical Considerations}
The presence of an object in the LoS neighborhood, as shown in Fig.~\ref{fig:geometry}, inevitably causes some of the propagating waves to reflect and scatter from its surface. The relative importance of reflection and scattering on the received signal depends on the wavelength and object's position, surface and electrical properties. These parameters define roughness of the surface which is the ratio of reflected and scattered RF power \cite{Pinel2010}. If the object does not have sharp edges compared to the wavelength, the ratio of mean fluctuating power of the electric field (incoherent power) to the power of mean electric field (coherent power) can be used for evaluating the roughness. Based on this ratio, the surfaces can be classified as \emph{flat}, \emph{slightly rough} or \emph{very rough}. For carrier frequencies considered in this paper, the average human body surface can be taken as slightly rough so that the flat surface reflection is the dominating effect altering the received signal. 

For the general reflection problem, the ratio between electric field intensities of incident and reflected waves is given by the Fresnel reflection coefficient \cite[ch. 4]{Rappaport2002}. The coefficient depends on polarization of the incident wave with respect to the plane of incidence, incidence angle $\theta$, electrical properties of the reflection surface, and the wavelength. When the TX antenna is standing in the direction perpendicular to plane of incidence ($z$ axis in Fig.~\ref{fig:geometry}), the electric field intensity of the emitted wave is oriented along the same axis. In other words, the reflected waves only have perpendicular polarization, which yields negative Fresnel reflection coefficient for all incidence angles. In what follows, we represent the Fresnel reflection coefficient with non-negative reflection coefficient $\Gamma$ and the sign as $\pi$ radians additional phase.

The reflection coefficient is affected by electrical properties of the dielectric boundary. Considering humans, clothing is the dielectric boundary, for which the relative permittivity $\epsilon_r = 1.5 - 2$ in $2.4 \text{ -- } 2.5 \text{ GHz}$ \cite{Sankaralingam2010}. Such values imply small reflected RF power when the incidence angle is large. On the other hand, for (dry) human skin, which is the second reflection layer, $\epsilon_r = 38$ in the same frequency band \cite{TissueElectric}. This value of $\epsilon_r$ implies high reflected RF power for all incidence angles. Although the reflected RF power has complex relation with electrical properties of the dielectric boundaries, the impact on RSS is expected to be large. In the remainder of the paper we assume unknown but constant $\Gamma \triangleq \Gamma(\theta, \epsilon_r)$ satisfying $0.2 \le \Gamma \le 0.7$.  

\subsection{Received Signal Power Measurements}

In case the propagation environment is completely empty except the person, there are two multipath components impinging on the receiver antenna; the LoS signal and the single bounce reflected signal. The received complex baseband signal in such a scenario can be written as \cite[ch.~13]{Proakis2008} 
\begin{equation*}
	r(t) = \sum\limits_{i=1}^{2}{\alpha_i(t)e^{-j 2\pi f_c \tau_i}s(t-\tau_i)} +n(t), 
\end{equation*}
where $f_c$ is the center frequency, $\alpha_i(t)$ and $\tau_i$ are amplitude and phase of $i$\textsuperscript{th} multipath component in respective order, $n(t)$ is complex Gaussian zero-mean noise with two-sided power spectral density $N_0/2$, and $s(t)$ is the transmitted signal of duration $T_s$. If the coherent receiver is synchronized to the LoS signal, $\tau_1 = 0$ and for the systems with no inter symbol interference, $0 < \tau_2 \ll T_s$. If we further suppose that the propagation medium is constant for the time period of interest, the amplitudes of individual components are constant, and the received signal is given by
\begin{equation}\label{eq:received_signal}
	r(t) = \alpha_1s(t) + \alpha_2 e^{-j \phi}s(t-\tau_2) + n(t), 
\end{equation}
where $\phi = 2 \pi f_c \tau_2 = 2 \pi \Delta f_c / c_0$. For a quadrature modulation scheme with the same in-phase and quadrature shaping pulses and when $n(t)$ is mean ergodic, the received signal power calculated for a large time window $T \gg T_s$ is given by
\begin{equation}\label{eq:received_signal_power}
	\mathcal{P}_r = \Big(\alpha_1^2 + \alpha_2^2 + 2 \alpha_1 \alpha_2 \cos(\phi)\Big) \mathcal{P}_s  + \hat{n},
\end{equation}
where $\hat{n}$ denotes the noise contaminating the power measurements.

The amplitude of the reflected signal is $\Gamma$ fraction of the incident signal, and it has $\pi$ radians additional phase due sign of the Fresnel reflection coefficient. Furthermore, the amplitude of both components in Eq.\eqref{eq:received_signal} have the same amplitude decay rate if both experience the same realization of fading, which can be modeled as a cite-dependent path-loss exponent $\eta$. In this case, the amplitudes of the components decrease with $(-\eta/2)^{\text{th}}$ exponent of the traversed path-length,
\begin{equation*} 
		\alpha_1 \approx  d^{-\eta/2}, \qquad
		\alpha_2 \approx -\Gamma \left(d+\Delta \right)^{-\eta/2},
\end{equation*}
where $\Delta$ is the excess path length traversed by the reflected waves. Substituting these into Eq.\eqref{eq:received_signal_power}  yields
\begin{equation} \label{eq:received_power}
\begin{aligned}
	\mathcal{P}_r = \frac{1+\Gamma^2 (1+{\Delta}/{d})^{-\eta}}{d^{\eta}}
				\Big(1 - \kappa(\Delta) \cos(\phi)\Big) \mathcal{P}_s  + \hat{n},
\end{aligned}
\end{equation}
where the amplitude of the sinusoidal term is
\begin{equation}\label{eq:kappa_Delta}
	\kappa(\Delta) \triangleq 2\Gamma\frac{(1+\Delta/d)^{\frac{\eta}{2}}}{\Gamma^2  + (1+\Delta/d)^\eta}.
\end{equation}

The derivations for the ideal scenario can be easily extended to cover more general conditions by including additional losses encountered in practical deployments. In logarithmic scale, these effects can be represented by a frequency dependent parameter $\mathcal{P}_0^{dB}(f_c)$ and additional stochastic parameters can be added to measurement noise. Therefore, the RSS measurement in logarithmic-scale can be written as 
\begin{equation}\label{eq:received_power_db1}
\mathcal{P}_r^{dB} = \mathcal{P}_{LoS}^{dB}(f_c) + \zeta(\Delta, \Gamma, f_c) + \nu(f_c), 
\end{equation} 
where $\nu(f_c)$ is the measurement noise and $\mathcal{P}_{LoS}^{dB}(f_c) = \mathcal{P}_0^{dB}(f_c)-10 \eta \log_{10}(d)$ is the LoS signal power. 

The LoS signal power $\mathcal{P}_{LoS}^{dB}(f_c)$ represents time invariant (or very slowly varying) properties of the measurements setup. If it is assumed constant throughout the experiment, it can be estimated when the environment is vacant. In this case, since $\nu(f_c)$ can be assumed zero mean and mean ergodic, an ensemble average of $K$ time samples is an estimate of $\mathcal{P}_{LoS}^{dB}(f_c)$,  
\begin{equation}\label{eq:pl_mle}
	\hat{\mathcal{P}}_{LoS}^{dB}(f_c) \triangleq \frac{1}{K}\sum\limits_{k=0}^{K-1} {\mathcal{P}_r^{dB}(k C T_p, f_c)},
\end{equation} 
where $k$ is the time sample index and $T_p$ is the packet rate at the channel with carrier frequency $f_c$. 
In the remainder of this paper, we suppose that the estimate $\hat{\mathcal{P}}_{Los}^{dB}(K, f_c)$ is subtracted from the measurements so that we have measurement model in the form
\begin{equation}\label{eq:power_measurement_dB}
	{z}(\Delta, \Gamma, f_c) \triangleq \zeta(\Delta, \Gamma, f_c) + \nu(f_c).
\end{equation}
Therefore, when the person is in the environment, $\zeta(\Delta, \Gamma, f_c)$ can be detected using signal in noise techniques. 

The effect of reflected signal in Eq.~\eqref{eq:power_measurement_dB} is given by  %
\begin{equation}\label{eq:effect_of_reflection_db}
\begin{aligned}
	\zeta(\Delta, \Gamma, f_c) \triangleq &~~~ 10 \log_{10} \bigg(1 + \Gamma^2 \Big({1+\Delta/d}\Big)^{-\eta}\bigg) \\
			&+10 \log_{10} \Big(1 -\kappa(\Delta) \cos(\phi) \Big).
\end{aligned}
\end{equation}
Since $\zeta(\Delta, \Gamma, f_c)$ is a periodic function, in Appendix~\ref{appendix:fourier_series} it is shown that 
\begin{equation}\label{eq:effect_of_reflection_db_series}
	\zeta(\Delta, \Gamma, \beta={f_{c}}/{c_0}) = -2 \hat{e} \sum\limits_{i=1}^{\infty} {a_i  \cos(2 \pi i \Delta \beta )},
\end{equation}
where 
\begin{equation}\label{eq:a_coefficients}
	a_i \triangleq {A^i}/{i},  \qquad A \triangleq {\Gamma}{(1+{\Delta}/{d})^{-\frac{\eta}{2}}}.
\end{equation}

It is to be noted that $\zeta(\Delta, \Gamma, f_c)$ in linear scale is a single tone shifted sinusoidal with frequency $\Delta$. However, inverting the measurement ${z}(\Delta, \Gamma, f_c)$ to linear scale changes the model and nature of the noise from additive to multiplicative. On the other hand, in logarithmic scale the reflected signal has periodic variations with communication frequency whose harmonics are non-zero for finite $\Delta$. Therefore, the received signal power measurements in logarithmic scale at different frequency channels are samples from a periodic function, whose frequency $\Delta$ also affects its Fourier coefficients $a_i$.


\section{Detection of Reflections} \label{sec:detection}
In this section, the probability of detection for a given false alarm probability is introduced. For this purpose, first the signal energy model and its two approximations are derived. Then, exact expressions for probability of false alarm and detection are given. Finally, numerical evaluations of the derivations in previous and this section are presented. 

\subsection{Signal Energy}
Let us denote the operation frequencies of the communication system by an ordered set $\mathscr{C} = \{{f_{c,l}}: f_{c,l+1} > f_{c,l}, l=1,2 \cdots C\}$, so that we have
\begin{equation*}
	\phi_l = 2 \pi \Delta\frac{f_{c,l}}{c_0} = 2 \pi \Delta\beta_l = 2 \pi \Delta \left(\beta_1 + (l-1) \xi \right).
\end{equation*} 
The total energy of the measurements given in Eq.~\eqref{eq:power_measurement_dB} for fixed $\Delta$ calculated over all the channels in $\mathscr{C}$ is given by
\begin{equation*}
	\mathcal{E}_z(\Delta, \Gamma) \triangleq \sum\limits_{l=1}^{C} \Big( \zeta(\Delta, \Gamma, \beta_l) + \nu_{l}\Big)^2,
\end{equation*}
where $\nu_{l} = \nu(\beta_{l})$. If zero-mean $\nu_l$ is also mean ergodic, energy of the measurements can be approximated by 
\begin{equation}\label{eq:energy_detector_model}
	\mathcal{E}_z(\Delta, \Gamma) \approx \mathcal{E}_{\zeta}(\Delta, \Gamma) + \sum\limits_{l=1}^{C} \nu_{l}^2,
\end{equation}
where $\mathcal{E}_{\zeta}(\Delta, \Gamma)$ is the energy of the signal in Eq.~\eqref{eq:effect_of_reflection_db}. The approximation improves as $C$ increases.

The signal in Eq.~\eqref{eq:effect_of_reflection_db} is a periodic function, which has Fourier series expansion with simple coefficients given in Eq.~\eqref{eq:a_coefficients}. Thus, the signal power can be written using Parseval's identity as 
\begin{equation}\label{eq:signal_power}
	\mathcal{P}_\zeta(\Delta, \Gamma) \triangleq 4 \hat{e}^2 \sum\limits_{i=1}^{\infty} a_i^2 = 
					4 \hat{e}^2 \sum\limits_{i=1}^{\infty} \frac{A^{2i}}{i^2} = 4 \hat{e}^2 \dilog(A^2),
\end{equation}
where $\dilog(\cdot)$ is the \emph{dilog} function \cite{Maximon2003}. The monotonicity of dilog function implies that the signal power increases with increasing $A$. In turn, $A$ is monotonically increasing function of $\Gamma$ and decreasing function of $\Delta$ so that the signal power is maximum for $\Delta = 0$. Furthermore, for $\Gamma \le 0.7$, the first two harmonics contain more than $96.76\%$ of the total signal power for all $\Delta$. This implies that for most practical purposes $\zeta(\Delta, \Gamma,\beta)$ can be approximated by 
\begin{equation}\label{eq:approximate_signal}
	 \bar{\zeta}(\Delta, \Gamma,\beta)  \triangleq -2 \hat{e} \Big( a_1 \cos(2\pi \Delta \beta) + a_2 \cos(4\pi \Delta \beta)\Big).
\end{equation} 
\begin{figure*}[!ht]
\normalsize
\begin{equation} \label{eq:reflected_energy}
\begin{aligned}
\mathcal{E}_{\bar{\zeta}} (\Delta, \Gamma) = 4 \hat{e}^2\Bigg( C \frac{a_1^2 + a_2^2}{2}
		 &+ a_1 a_2\cos(2\pi \Delta \bar{\beta})\frac{\sin(\pi \Delta C \xi)}{\sin(\pi \Delta \xi)} 
		 + \frac{a_1^2}{2}\cos(4\pi \Delta \bar{\beta})\frac{\sin(2\pi \Delta C \xi)}{\sin(2\pi \Delta \xi)} \\
		 &+ a_1 a_2\cos(6\pi \Delta \bar{\beta})\frac{\sin(3\pi \Delta C \xi)}{\sin(3\pi \Delta \xi)} 
		 + \frac{a_2^2}{2}\cos(8\pi \Delta \bar{\beta})\frac{\sin(4\pi \Delta C \xi)}{\sin(4\pi \Delta \xi)} \Bigg)
\end{aligned}
\end{equation}
\hrulefill
\vspace*{1pt}
\end{figure*} 

In Appendix~\ref{appendix:reflected_signal_energy}, it is shown that the energy of $\bar{\zeta}(\Delta, \Gamma, \beta)$, denoted by $\mathcal{E}_{\bar{\zeta}}(\Delta, \Gamma)$, for odd $C$ is given by Eq.~\eqref{eq:reflected_energy}, and the exact expression for even $C$ is more involved but it can be approximated using Eq.~\eqref{eq:reflected_energy}. 

Accuracy of the approximating $\mathcal{E}_{{\zeta}}(\Delta, \Gamma)$ with $\mathcal{E}_{\bar{\zeta}}(\Delta, \Gamma)$ depends on $\Delta$ and amplitude of the higher order harmonics. For large $\Delta$ $\big(\Delta > 1\big/(2 C \xi)\big)$, the difference between $\mathcal{E}_{\bar{\zeta}}(\Delta, \Gamma)$ and $\mathcal{E}_{\zeta}(\Delta, \Gamma)$ is low since the measurements contain frequencies more than one period of the second and higher order harmonics. In this $\Delta$ region, the signal energy approaches to its average value and it is possible to further approximate $\mathcal{E}_{{\zeta}}(\Delta, \Gamma)$ with the first term of Eq.~\eqref{eq:reflected_energy}, that is,
\begin{equation}\label{eq:reflected_energy_approx2}
	\bar{\mathcal{E}}_{\bar{\zeta}}(\Delta, \Gamma) \triangleq 4 \hat{e}^2 C \frac{a_1^2 + a_2^2}{2}  
		\approx \mathcal{E}_{\bar{\zeta}}(\Delta, \Gamma)\approx \mathcal{E}_{{\zeta}}(\Delta, \Gamma).
\end{equation}

\subsection{Energy Detector}
In this paper, we are interested in detecting the reflected signal, which can be formulated as binary hypothesis testing. The measurements given in Eq.~\eqref{eq:energy_detector_model} can be tested for presence of the reflected signal against absence of it, that is,
\begin{equation*}
\begin{aligned}
	 \mathcal{H}_0: \mathcal{E}_z &=  \sum\limits_{l=1}^{C}\nu_{l}^2, \qquad
	 \mathcal{H}_1:\mathcal{E}_z &=   \mathcal{E}_\zeta +  \sum\limits_{l=1}^{C}\nu_{l}^2
\end{aligned}
\end{equation*}
An energy detector makes a decision based on a threshold associated with the decision criteria of the binary hypothesis testing \cite[ch. 2]{Trees1968}. If priori probabilities of the hypothesis are known and it is possible to assign costs for transitions, then the \emph{Bayesian} criterion can be utilized. If either of these cannot be determined, the \emph{Neyman-Pearson} criterion should be used. For the considered problem it is not an easy task to determine prior probabilities and assign costs for transitions, thus the \emph{Neyman-Pearson} criterion is better suited. 
   
Let us denote the probability of false alarm by $P_f \triangleq \Pr\left\{ \mathcal{E}_z > X \big| \mathcal{H}_0\right\}$ and probability of detection by $P_d \triangleq \Pr\left\{ \mathcal{E}_z > X \big| \mathcal{H}_1\right\}$ for decision threshold $X$. The \emph{Neyman-Pearson} criterion is to constrain $P_f$ and then seek for a decision threshold that maximizes $P_d$. Therefore, performance of a detector can be evaluated by variation of $P_d$ with $P_f$, which is also known as the \emph{receiver operating characteristic} (ROC).

In case the Fourier coefficients in Eq.~\eqref{eq:a_coefficients} are taken as deterministic, the problem at hand is energy detection of an unknown deterministic signal. If it is further supposed that $\nu_{l}$ is an independent and identically distributed (i.i.d.) sequence of Gaussian random variables, the problem corresponds to the one studied by Urkowitz in his seminal work on the subject \cite{Urkowitz1967}. The aforementioned problem is extended for quasi-deterministic signals by Kostylev \cite{Kostylev2000}, and Digham \emph{et al.} has given closed form expressions for probability of detection and false alarm \cite{Digham2007}. The energy detection problem for i.i.d. noise sequence case has also been studied for random amplitudes \cite{Digham2007}, but we assume that the unknown signal is deterministic and do not consider statistical properties of the Fourier coefficients. 

In Appendix \ref{appendix:quadratic_characteristic_equation}, it shown that the distribution of the energy detector output is tractable when $\nu_{l}$ is i.i.d. or  when means are different but variances are identical. Energy detector output for other cases has different characteristics such that the corresponding distribution function either does not have a closed form expression or only has an infinite series representation. Therefore, we give $P_d$ and $P_f$ for i.i.d. $\nu_{l}$.

If $\mathcal{H}_0$ is true, the output of the energy detector is the square sum of $C$ zero-mean Gaussian random variables with equal variance $\sigma^2$. For this case, $\mathcal{E}_z(\Delta, \Gamma)$ has $C$ degrees of freedom chi-squared distribution with density function \cite[p. 45]{Proakis2008}
\begin{equation}
	f_\chi(\mathcal{E}_z) = \frac{1}{\left( 2 \sigma^2 \right)^{C/2} \gamma(C/2, 0)} \mathcal{E}_z^{\frac{C}{2}-1} e^{-\frac{\mathcal{E}_z}{2 \sigma^2}},
\end{equation}
where $\gamma(\cdot, \cdot)$ is (upper) \emph{incomplete gamma} function \cite[sec. 6.5]{Abramowitz1970}. On the other hand, in case $\mathcal{H}_1$ is true, $\mathcal{E}_z(\Delta, \Gamma)$ has $C$ degrees of freedom non-central chi-squared distribution with non-centrality parameter $\mu$, whose density function is given by \cite[p. 46]{Proakis2008}
\begin{equation}\label{eq:non_central_chi_square}
	f_{\chi'}(\mathcal{E}_z) = \frac{\left(\frac{\mathcal{E}_z}{\mu} \right)^{\frac{C-2}{4}}}{2 \sigma^2}  e^{-\frac{\mathcal{E}_z+\mu}{2 \sigma^2}}
				   I_{\frac{C}{2}-1}\left(\frac{\sqrt{\mu \mathcal{E}_z}}{\sigma^2}\right),
\end{equation} 
where $I_v(\cdot)$ is $v^{\text{th}}$ order modified Bessel function of first kind \cite[sec. 9.6]{Abramowitz1970}. The non-centrality parameter of the distribution in Eq.~\eqref{eq:non_central_chi_square} is $\mu = {\mathcal{E}_\zeta(\Delta, \Gamma)}$. For these density functions of $\mathcal{E}_z$, $P_f$ and $P_d$ are given by
\begin{align}
	P_f &= \bar{\gamma}\left(\frac{C}{2}, \frac{X}{2 \sigma^2}\right) 
		 = \frac{\gamma \left(\frac{C}{2}, \frac{X}{2 \sigma^2}\right)}{\gamma \left(\frac{C}{2}, 0 \right)} \label{eq:false_alarm}\\  
	P_d &= Q_{C/2}\left( \sqrt{\frac{\mathcal{E}_\zeta}{\sigma^2}}, \sqrt{\frac{X}{\sigma^2}}\right) \label{eq:detection_probability}
\end{align}
where $\bar{\gamma}(\cdot,\cdot)$ is regularized upper incomplete gamma function, and $Q_{C/2}(\cdot, \cdot)$ is the generalized \emph{Marcum Q-}function. It is important to note that the expression of $P_d$ is only valid for even $C$. 

Since $\bar{\gamma}(\cdot,\cdot)$ has an inverse $\bar{\gamma}^{-1}(\cdot,\cdot)$, the decision threshold can be determined from $P_f$ using
\begin{equation}\label{eq:threshold}
	X = 2 \sigma^2 \bar{\gamma}^{-1}\left(\frac{C}{2},P_f\right).
\end{equation}  
Therefore, ROC is obtained as a function of $\Delta$, $\Gamma$ and $d$ by writing $P_d$ in terms of $P_f$ as 
\begin{equation}\label{eq:roc}
	P_d = Q_{C/2}\left( \sqrt{\frac{\mathcal{E}_\zeta}{\sigma^2}}, \sqrt{2 \bar{\gamma}^{-1}\left(\frac{C}{2},P_f\right)}\right).
\end{equation}  

\begin{figure*}[ht]
\centering
\begin{tabular}{C{5.65cm}C{5.65cm}C{5.65cm}}
\subfloat[]{\includegraphics[width=5.65cm]{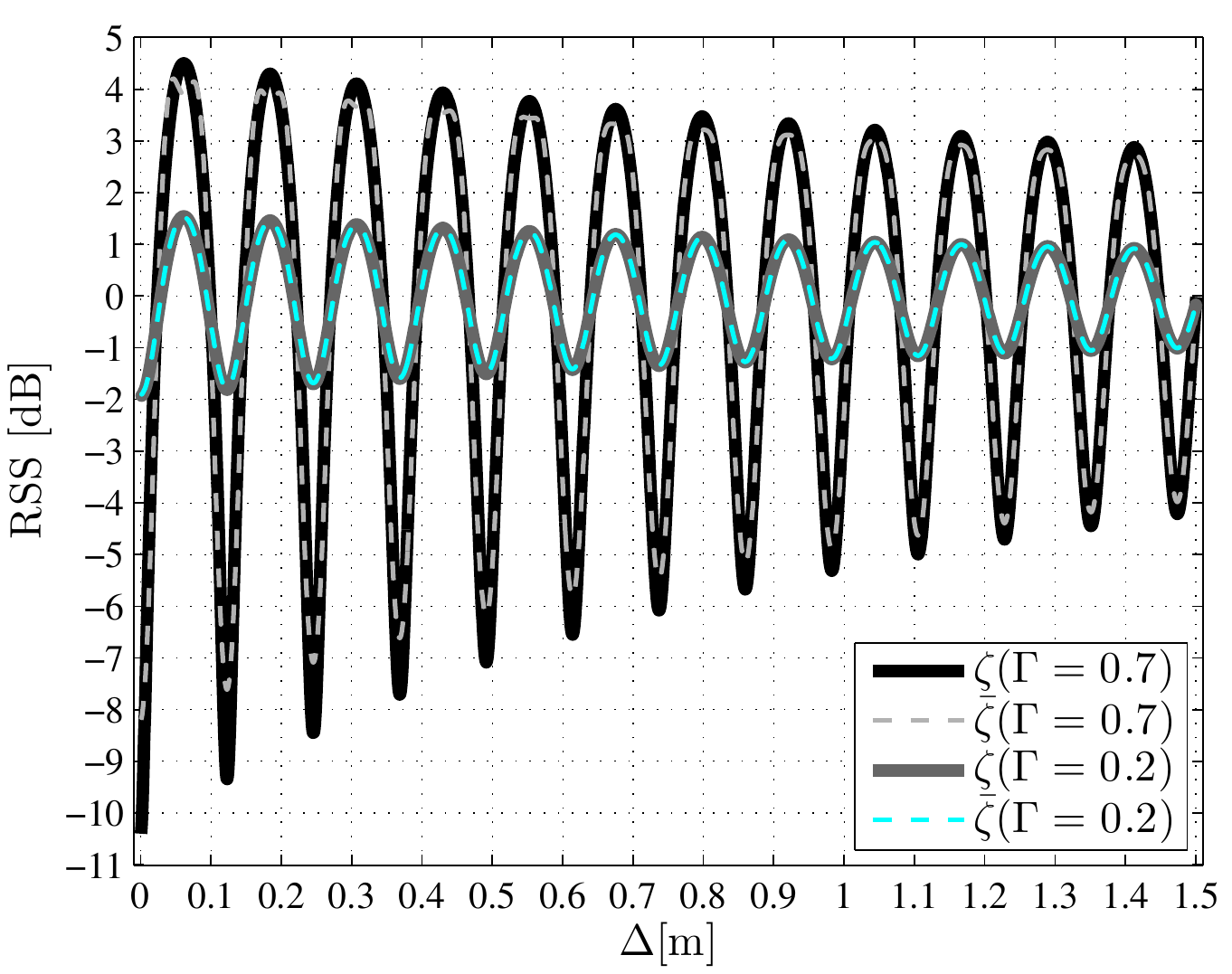}\label{fig:zeta_Delta}}&
\subfloat[]{\includegraphics[width=5.65cm]{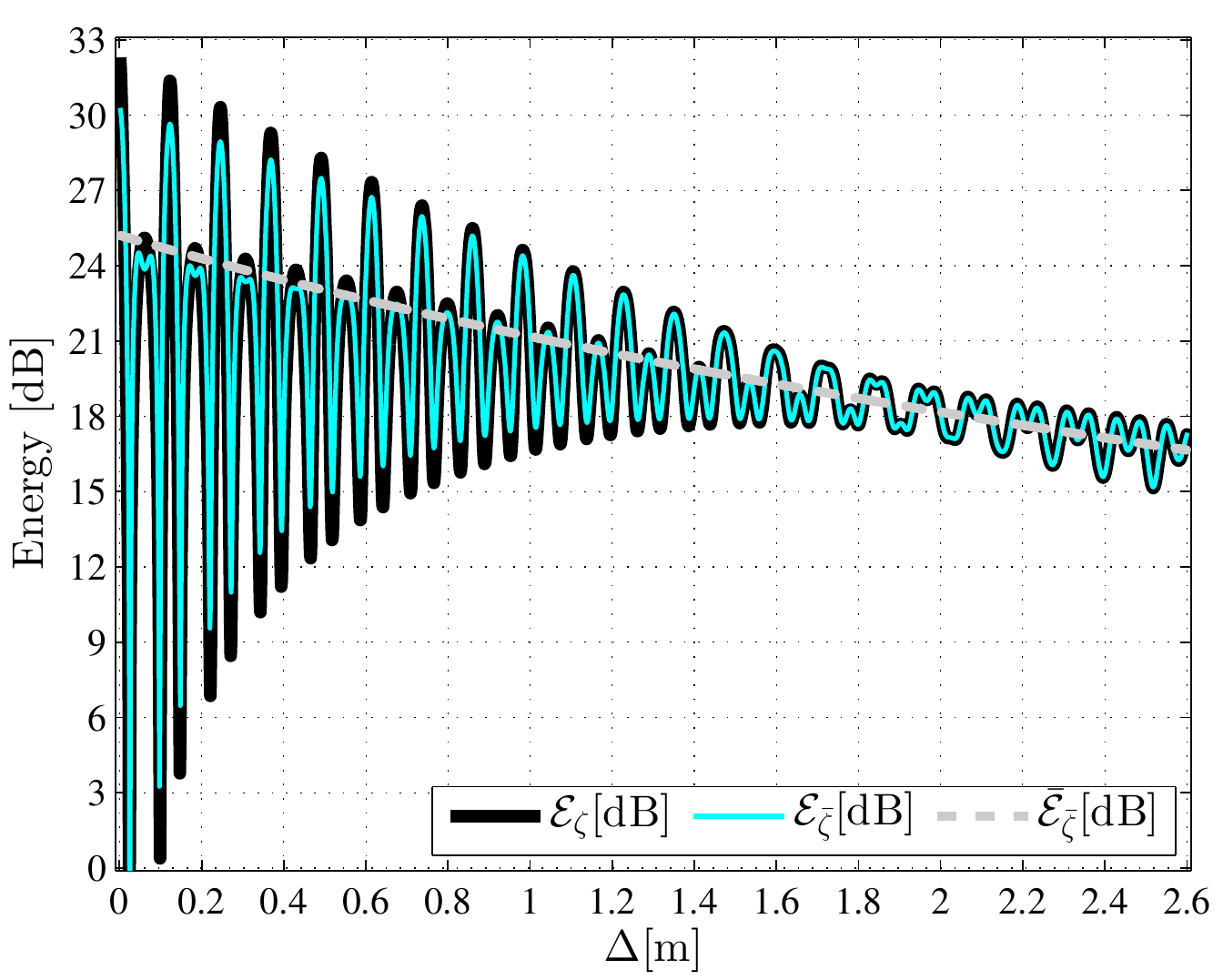}\label{fig:energy_delta}} & 
\subfloat[]{\includegraphics[width=5.65cm]{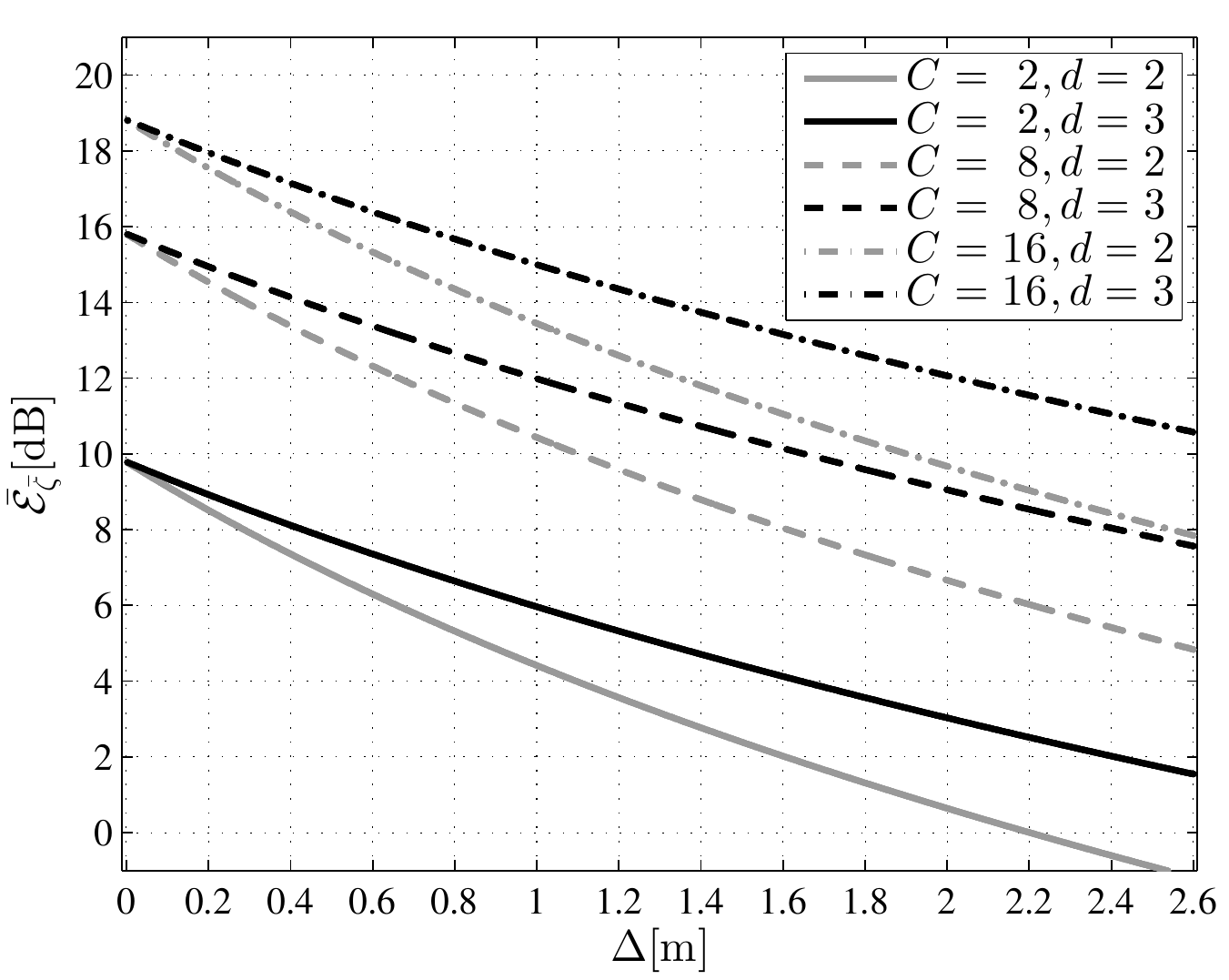}\label{fig:energy_xi_c_d}}
\end{tabular}
\caption{Numerical evaluation of reflected signal. In (a), variation of $\zeta(\Delta, \Gamma, \beta)$ and $\bar{\zeta}(\Delta, \Gamma, \beta)$ with $\Delta$ and $\Gamma \in\{0.2, 0.7\}$ at $2.4425 \text{ GHz}$ for $d = 3 \text{ m}$. In (b), variation of $\mathcal{E}_\zeta(\Delta, \Gamma)$, $\mathcal{E}_{\bar{\zeta}}(\Delta, \Gamma)$ and $\bar{\mathcal{E}}_{\bar{\zeta}}(\Delta, \Gamma)$ in dB with $\Delta$ for $\Gamma = 0.7$, $C=16$ and $d = 3 \text{ m}$. In (c), variation of $\bar{\mathcal{E}}_{\bar{\zeta}}(\Delta, \Gamma)$ with $\Delta$,  $d \in \{2,3\} \text{ m}$ and $C \in \{2, 8, 16\}$ for $\Gamma=0.35$.}
\label{fig:numerical_results1}

\centering
\begin{tabular}{C{5.65cm}C{5.65cm}C{5.65cm}}
\subfloat[]{\includegraphics[width=5.65cm]{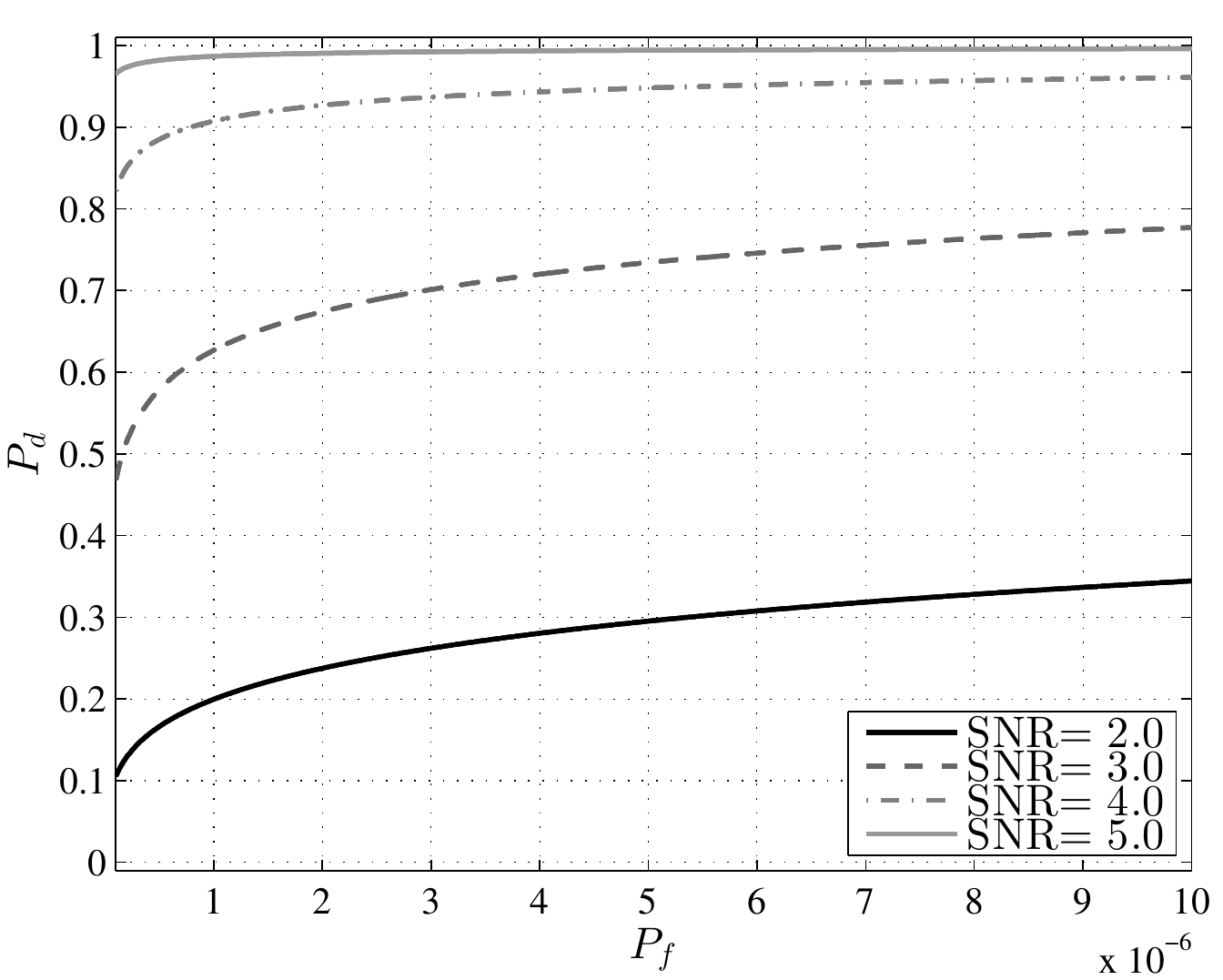}\label{fig:roc}}&
\subfloat[]{\includegraphics[width=5.65cm]{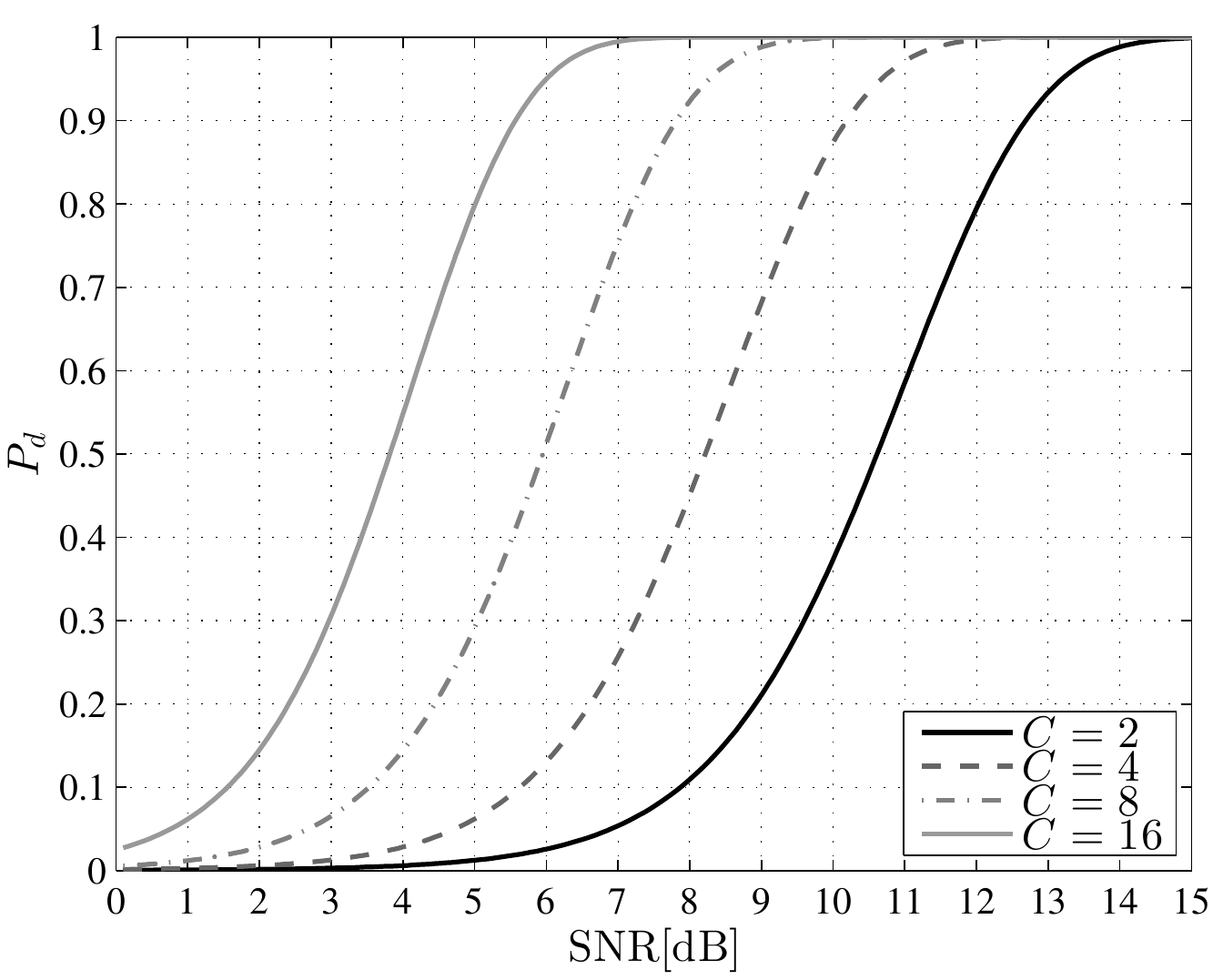}\label{fig:pd_c}} & 
\subfloat[]{\includegraphics[width=5.65cm]{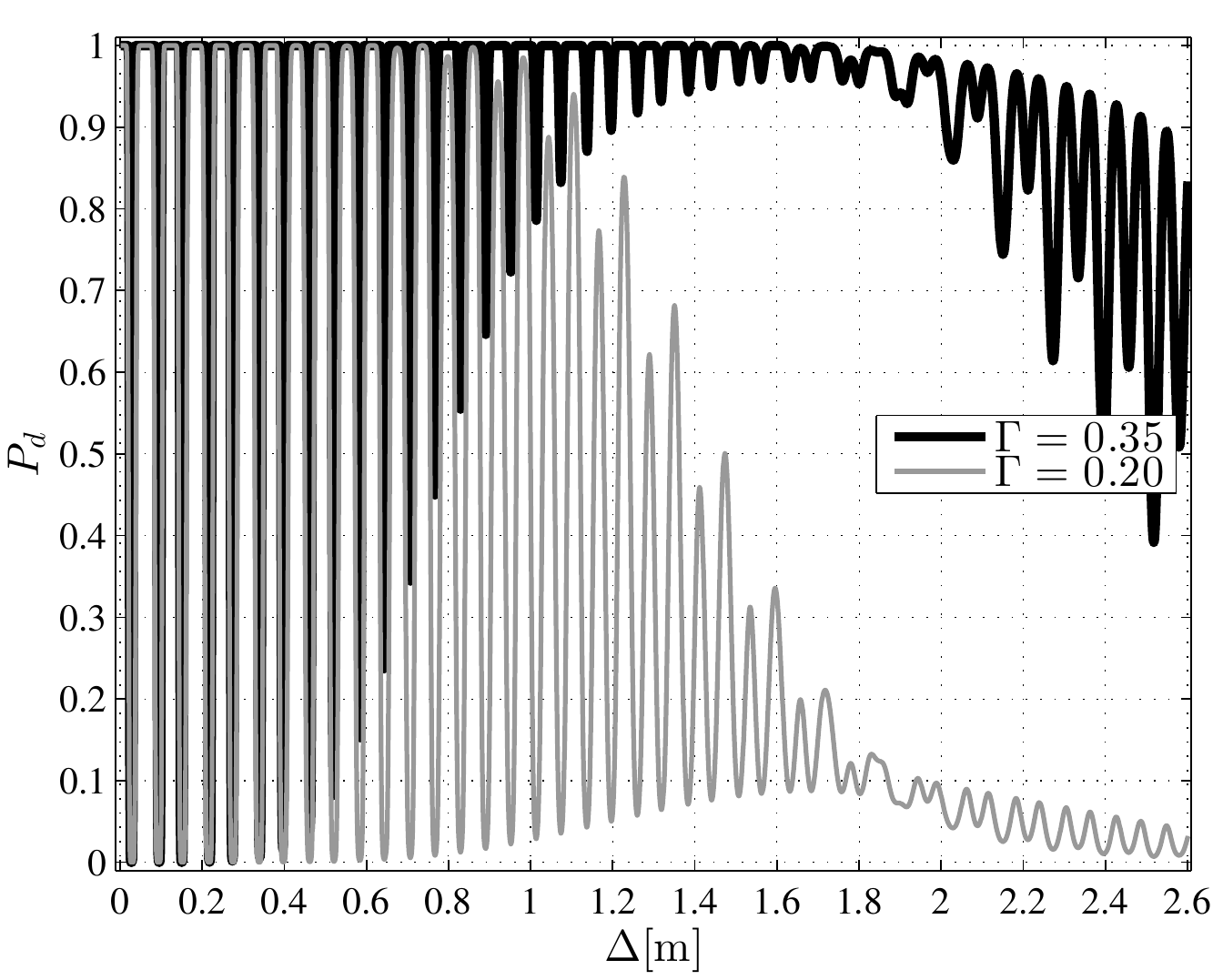}\label{fig:pd_Delta}}
\end{tabular}
\caption{Numerical evaluation of probability of detection. In (a), ROC for different $\mathcal{E}_{\zeta}\big/(C \sigma^2)$ for $C=16$. In (b), variation of $P_d$ with $\mathcal{E}_{\zeta}\big/(C \sigma^2)$ in dB and $C\in\{2,4,8,16\}$ for $P_f = 6 \cdot 10^{-6}$. In (c),  variation of $P_d$ with $\Delta$ and $\Gamma \in \{0.20,0.35\}$ for $d = 3 \text{ m}$, $P_f = 6 \cdot 10^{-6}$, $C=16$, and $\sigma=0.5$.}
\label{fig:numerical_results2}
\end{figure*} 
\subsection{Numerical Evaluation} \label{sec:numerical_results}

The derivations given up to this point are evaluated numerically and the results are shown in Fig.~\ref{fig:numerical_results1} and Fig.~\ref{fig:numerical_results2}. In Fig.~\ref{fig:numerical_results1}, the reflected signal and its energy derivations are evaluated using the numerical values calculated for $c_0=3\cdot10^8 \text{ m/s}$ and $\eta = 3$ over frequency band defined by $2.405 \text{ GHz}$ starting frequency and $80 \text{ MHz}$ bandwidth. The center frequency of the channels are equally spaced and the center frequency of the band is fixed to $2.4425 \text{ GHz}$. In Fig.~\ref{fig:numerical_results2}, probability of detection related numerical results are shown.

The validity of Fourier series expansion in Eq.~\eqref{eq:effect_of_reflection_db} is evaluated through its two term approximation given in Eq.~\eqref{eq:approximate_signal}. In Fig.~\ref{fig:zeta_Delta}, the variation of Eq.~\eqref{eq:effect_of_reflection_db} and Eq.~\eqref{eq:approximate_signal} with $\Delta$ for $\Gamma=0.2$ and $\Gamma=0.7$ at $f_c = 2.4425 \text{ GHz}$ are shown. For $\Gamma=0.2$, there is no visible difference between the reflected signal and its approximation.  However, once $\Gamma$ is increased to $0.7$ there is a slight observable difference. This follows from the fact that the signal power given in Eq.~\eqref{eq:signal_power} and Fourier series coefficients in Eq.~\eqref{eq:a_coefficients} are monotonically increasing functions of $\Gamma$, and higher order harmonics increase faster than the first two harmonics. However, the approximation is accurate enough for the cases satisfying the prior assumption $\Gamma \le 0.7$.

The reflected signal energy approximations given in Eq.~\eqref{eq:reflected_energy} and Eq.~\eqref{eq:reflected_energy_approx2} are validated by calculating their variation with $\Delta$ for $\Gamma=0.7$ as shown in Fig.~\ref{fig:energy_delta}. The numerical values are calculated for $16$ of $5 \text{ MHz}$ separated channels in the specified band, which correspond to frequency channels of $2.4 \text{ GHz}$ ISM band IEEE 802.15.4-2006 radios. As expected, the reflected signal energy decreases with increasing $\Delta$. Furthermore, the approximation given in Eq.~\eqref{eq:reflected_energy} is very close to the true signal energy for all values of $\Delta$. On the other hand, the approximation given in Eq.~\eqref{eq:reflected_energy_approx2} is close to the energy for $\Delta > 1.5$, which is in coherence with the value $1 \big/ (2 C \xi) = 1.875 \text{ m}$ and the energy is close to its mean value. To note, the difference between ${\mathcal{E}}_{{\zeta}}(\Delta, \Gamma)$, ${\mathcal{E}}_{\bar{\zeta}}(\Delta, \Gamma)$ and $\bar{\mathcal{E}}_{\bar{\zeta}}(\Delta, \Gamma)$ is the highest at $\Delta = l \big/\bar{\beta} = 0.1228 l \text{ m}$ for $l=1,\cdots$. Therefore, the probability of detection is not expected to have uniform spatial distribution.

The average reflected signal energy $\bar{\mathcal{E}}_{\bar{\zeta}}(\Delta, \Gamma)$, defined in Eq.~\eqref{eq:reflected_energy_approx2}, as a function of $\Delta$, for 
 $C\in\{2, 8, 16\}$, and $d\in\{2, 3\} \text{ m}$ are depicted in Fig.~\ref{fig:energy_xi_c_d}. As shown, $\bar{\mathcal{E}}_{\bar{\zeta}}(\Delta, \Gamma)$ grows with increasing $d$ as implied by definition of $A$ in Eq.~\eqref{eq:a_coefficients}. Furthermore,  $\bar{\mathcal{E}}_{\bar{z}}(\Delta, \Gamma)$ is an increasing function of $C$. However, the rate of increase is decreasing, that is, the gain is the most significant for small number of frequency channels. Therefore, for a given band it is possible to find an optimal number of channels that yields a desired reflected signal energy on the average.

  
In Fig.~\ref{fig:roc}, the ROC of the energy detector for $C=16$ and different signal-to-noise ratio (SNR) values, given by
\begin{equation*}
	SNR \triangleq {\mathcal{E}_\zeta}\big/{\left(C \sigma^2\right)},
\end{equation*}
are shown. As the SNR increases, the probability of detection increases for $P_f \in [ 1\cdot 10^{-7}, 1\cdot10^{-5} ]$ as implied by Eq.~\eqref{eq:detection_probability}. For $C=16$, $P_f = 6\cdot10^{-6}$ yields high detection probability when the SNR is $4$ or more. For this particular $P_f$ value (corresponding to threshold $X=53.6117 \sigma^2$ for $C=16$), variation of $P_d$ with SNR and $C\in\{2, 4, 8, 16\}$ are shown in Fig.~\ref{fig:pd_c}. For a fixed SNR, the detection probability increases with increasing number of channels because the approximation in Eq.~\eqref{eq:energy_detector_model} becomes more accurate. 

\begin{figure}[t]
\centerline{\includegraphics[width=0.9\columnwidth]{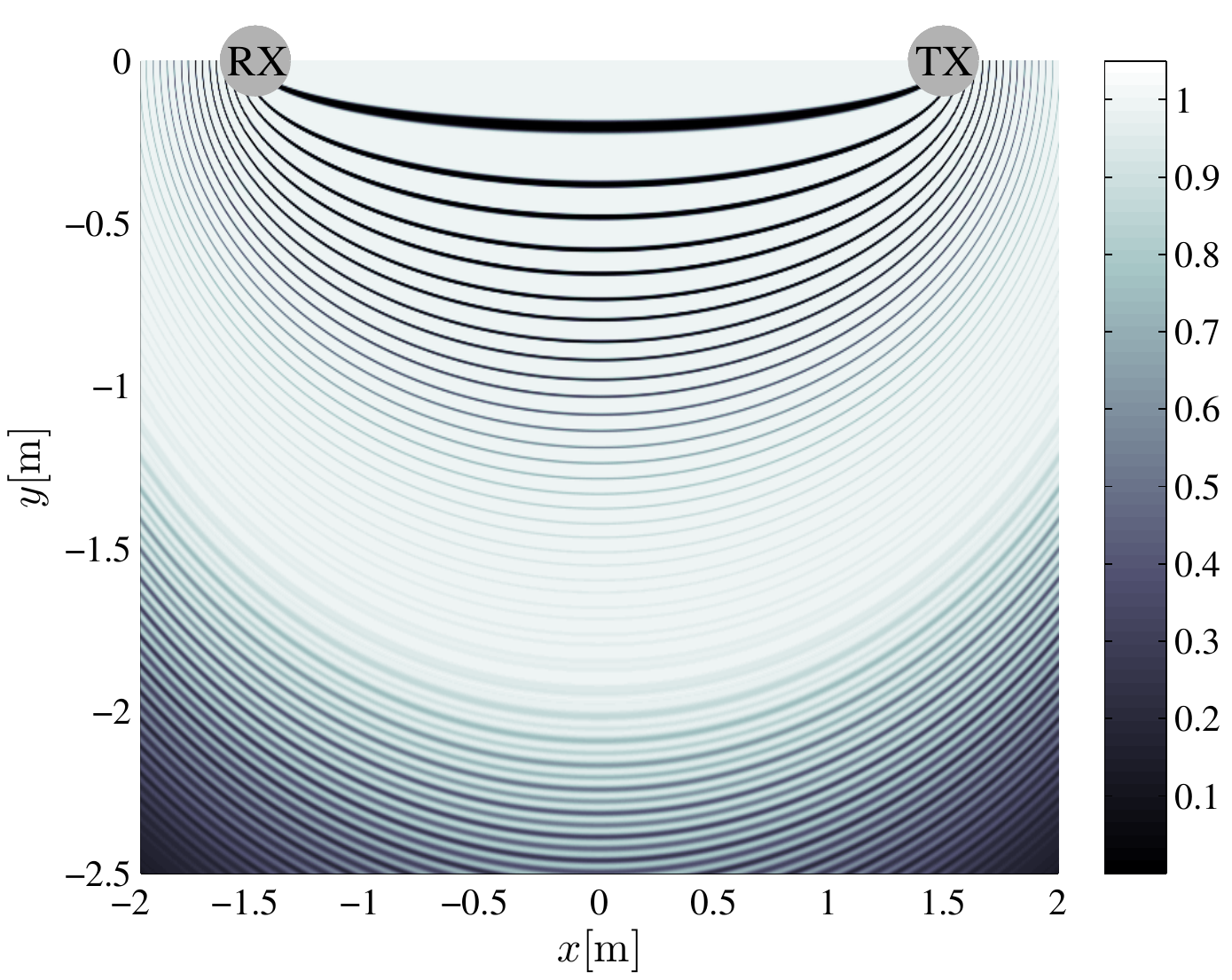}}
\caption{Spatial distribution of detection probability of a TX-RX pair with $d = 3 \text{ m}$ for $\Gamma = 0.35$, $P_f = 6 \cdot 10^{-6}$, $C=16$, and $\sigma=0.5$}
\label{fig:pd_region}
\end{figure}

In Fig.~\ref{fig:pd_Delta}, the variation of $P_d$ with $\Delta$ for $\Gamma \in \{0.2, 0.35\}$, $P_f = 6\cdot10^{-6}$, $C=16$ and $\sigma = 0.5$ is shown. Since the average reflected signal energy increase with $\Gamma$, on the average $P_d$ increases with $\Gamma$. However, the detection probability is not a monotonic function of $\Delta$. Rather, it has valleys especially when $\Delta$ is small ($ < 1.0 \text{ m}$) for both $\Gamma$ values. The spatial distribution of detection probability of the TX-RX pair evaluated in Fig.~\ref{fig:pd_Delta} is shown in Fig.~\ref{fig:pd_region} for $\Gamma = 0.35$. There are narrow regions close to LoS with zero detection probability. However, on the points yielding $1.5 < \Delta < 2$, the detection probability is high. Therefore, the energy detector can be used for detecting the reflected signal in a relatively large region around LoS with high probability even with low quality receivers provided that $\Gamma \ge 0.35$.

\section{Experimental Validation} \label{sec:experimental_validation}

Measuring the impact of human-induced reflections on RSS, in general, requires accurately locating the person since the excess path length is very sensitive to misplacement. The rate of change of excess path length with either of the position coordinates is a monotonically increasing function of the absolute value of the coordinates. In other words, when the person is not obstructing the LoS, a very small displacement causes the excess path length to change comparable to wavelength, which in turn changes impact of the reflected signal in accordance with Eq.~\eqref{eq:effect_of_reflection_db}. Therefore, evaluation of the presented development is a difficult task.

In this section, the empirical data acquired using the experimental setup depicted in Fig.~\ref{fig:experiment_setup} are used for showing that the presented energy detector can detect a person's presence in a relatively large area around the LoS. We first give the details of the measurement setup and conducted experiments, and then summarize the detection performance with the acquired measurements.

\subsection{Measurement Setup and Experiments}

\begin{figure}[t]
\centerline{\includegraphics[width=0.75\columnwidth]{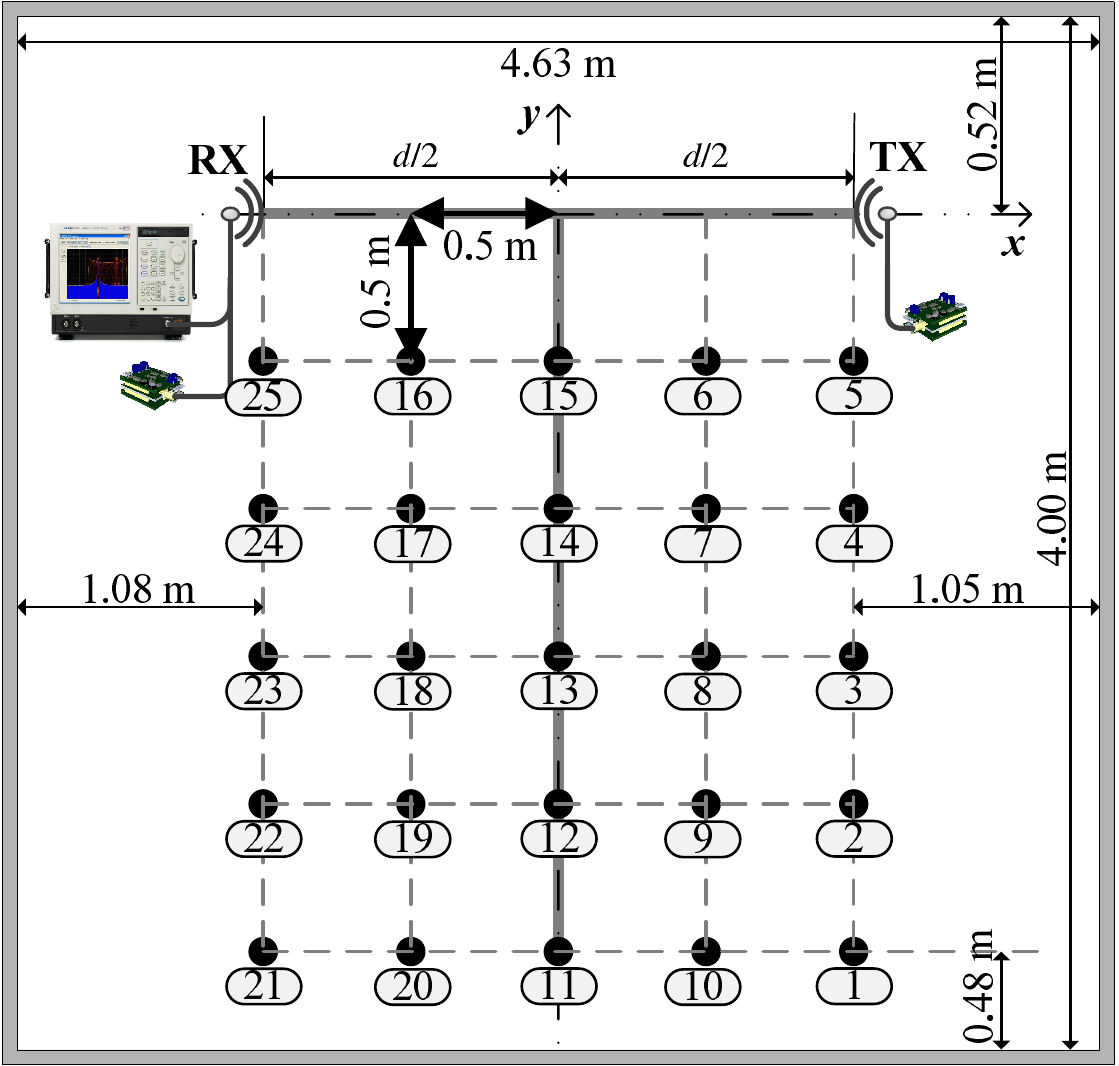}}
\caption{Experimental setup}
\label{fig:experiment_setup}
\end{figure}

The measurement setup illustrated in Fig.~\ref{fig:experiment_setup} is composed of a single antenna pair. One of the antennas is directly connected to the TX node while the other antenna is connected to a directional coupler to enable simultaneous acquisition of the RSS using a real-time spectrum analyzer (RSA) and the RX node. In order to restrict the reflections to the region of interest (the gridded region in Fig.~\ref{fig:experiment_setup}), vertically polarized directional antennas with $90^\circ$ horizontal beam width, almost flat gain pattern and $12 \text{ dBi}$ gain are utilized. The used nodes are equipped with Texas Instruments CC2431 IEEE 802.15.4 PHY/MAC compliant transceivers. The transceivers' micro-controller unit runs a communication software and a modified version of the FreeRTOS micro-kernel operating system. 


The TX node is programmed to transmit packets over each of the $16$ frequency channels defined by the IEEE 802.15.4-2006 standard at the $2.4$ GHz ISM band according to the schedule shown in Fig.~\ref{fig:schedule}. The transmission interval is aligned to $2 \text{ ms}$ corresponding to $T_p = 32 \text{ milliseconds}$ for each frequency channel. The RX node timestamps the packets with a resolution of $1/32$ microseconds when the start of frame delimiter is received. The packets are stored to a SD memory card of the node for off-line analysis. On average, the time interval between receptions is $2.08\text{ milliseconds}$ with a standard deviation of $160.00\text{ microseconds}$. The RSA is configured to cover the frequency band between $2.400 \text{ GHz}$ to $2.490 \text{ GHz}$ with $500 \text{ kHz}$ resolution and to acquire samples for a duration of $128 \text{ microseconds}$ in DPX mode of operation. The amplitude trace of the $90 \text{ MHz}$ bandwidth is represented using $10401$ points. The operation of the RSA is remotely controlled by a portable computer. A query is sent to the RSA to output the peak amplitude trace data calculated for the most recent dwell time. The result is read asynchronously on average once every $47.76 \text{ ms}$ with a standard deviation of $11.89 \text{ ms}$. The acquired trace values together with the time of request information are stored for off-line analysis.

The aforementioned measurement setup is placed in a $4.63 \times 4.00 \text{ m\textsuperscript{2}}$ office room. The room area is represented by a grid of $25$ points, which are $0.5 \text{ m}$ apart from each other in both spatial directions. A unique identifier is assigned to each grid point in the room as visualized in Fig.~\ref{fig:experiment_setup}. During the experiments, the person stands still for $30 \text{ seconds}$ on each grid point, starting from the point with identifier $1$. After $30 \text{ seconds}$, the person moves to the next point in the grid with $+1$ identifier. 
Before the person enters to the room, two empty room measurement sets are acquired (both for $2 \text{ minutes}$). One of the data sets is used for calculating $\hat{\mathcal{P}}_{LoS}^{dB}(f_c)$ and $\sigma^2$. The other is used for false alarm rate calculation. The experiments are repeated for $d = 2 \text{ and } 3 \text{ m}$. 

\subsection{Practical Considerations and Measurement Pre-processing}

The introduced detector is built upon the assumption that the measurement noise at the different frequency channels is an i.i.d. zero-mean Gaussian sequence. However, the probability density function of the measurement noise is typically a skewed function as earlier reported by several researchers \cite{Wilson2012, Zheng2012}. The acquired measurements of both systems also exhibit the same behavior, inevitably affecting the detection performance. In order to overcome this difficulty, the variance of the noise should be increased to reach a desired false alarm rate. This approach is equivalent to finding a Gaussian density which covers the tails of the skewed measurement noise density. Therefore, empty room measurements should be used for calibrating $\sigma^2$.

The RSS measurements of the RSA and RX node have different characteristics. In the aforementioned configuration, the RSA is operated as a wideband incoherent receiver which can output high resolution spectral measurements. Thus, the calculated RSS values are contaminated with wideband noise, which has the same moments for all frequency channels. On the other hand, low-resolution RSS measurements of the RX node are contaminated with narrowband noise, which can have different moments at different frequency channels. This difference can originate from small variations in frequency response of the RF front-end hardware. Therefore, the RSA measurements can be used as reference measurements.  

The acquired measurements are low-pass filtered by a $64^{\text{th}}$ order FIR filter in \emph{direct form II} with $1 \text{ Hz}$ cut-off frequency and maximally flat frequency response. The selected cut-off frequency is lower than the expected Doppler spread of a typical indoor channel. However, since the person is stationary, the filter removes out-of-band noise. More importantly, the low quality measurements of the RX node reaches the quality of the RSA measurements \cite{Kaltiokallio2014}. The increased response time of the detector, on the other hand, should be carefully evaluated for requirements of the deployment. The selected filter order is suitable for proof-of-concept purposes. As stated during the derivation of Eq.~\eqref{eq:power_measurement_dB}, $\hat{\mathcal{P}}_{LoS}^{dB}(f_c)$ are subtracted from the RSS measurements after filtering.  

\subsection{Results}

\begin{table}[ht]
\renewcommand{\arraystretch}{0.9}
\renewcommand{\tabcolsep}{2.2 mm}
\caption{Empirical Parameters} \label{table:constant_parameters}
\centering
\begin{tabular}[c]{l c c c c}
\toprule [2pt]
\textbf{Experiment}  & \textbf{Condition} & \textbf{Parameter} & \textbf{RSA} & \textbf{RX Node} \\
\midrule [1pt]
\multirow{14}{*}{$d=2 \text{ m}$} &  & $\sigma$ & $0.1328$ & $0.1325$\\
							& & $\hat{\sigma}$& $0.1328$ & $0.1600$ \\ \cmidrule[0.5pt]{2-5}
							& \multirow{2}{*}{$C=\;2$}	 & $X$		   & $0.4240$ & $0.6156$ \\
							&							 & $\hat{P}_f$ & $0.0015$ 		&$0.0000$	 \\\cmidrule[0.5pt]{2-5}
							& \multirow{2}{*}{$C=\;4$}	 & $X$		   & $0.5213$ & $0.7569$ \\
							&							 & $\hat{P}_f$ & $0.0008$ 		&$0.0000$	 \\\cmidrule[0.5pt]{2-5}
							& \multirow{2}{*}{$C=\;8$}	 & $X$		   & $0.6794$ & $0.9864$ \\
							&							 & $\hat{P}_f$ & $0.0023$ 	  &$0.0000$	 \\\cmidrule[0.5pt]{2-5}
							& \multirow{2}{*}{$C=16$}	 & $X$		   & $0.9453$ & $1.3725$ \\
							&							 & $\hat{P}_f$ & $0.0000$ 	  &$0.0017$		 \\
\midrule [1pt] 
\multirow{14}{*}{$d=3 \text{ m}$} &  & $\sigma$ & $0.1380$ & $0.1567$\\
							& & $\hat{\sigma}$& $0.1380$ & $0.1900$ \\ \cmidrule[0.5pt]{2-5}
							& \multirow{2}{*}{$C=\;2$}	 & $X$		   & $0.4580$ & $0.8681$ \\
							&							 & $\hat{P}_f$ & $0.0034$ 		&$0.0028$		 \\\cmidrule[0.5pt]{2-5}
							& \multirow{2}{*}{$C=\;4$}	 & $X$		   & $0.5631$ & $1.0673$ \\
							&							 & $\hat{P}_f$ & $0.0041$ 		&$0.0011$		 \\\cmidrule[0.5pt]{2-5}
							& \multirow{2}{*}{$C=\;8$}	 & $X$		   & $0.7339$ & $1.3910$ \\
							&							 & $\hat{P}_f$ & $0.0030$ 		&$0.0006$		 \\\cmidrule[0.5pt]{2-5}
							& \multirow{2}{*}{$C=16$}	 & $X$		   & $1.0212$ & $1.9354$ \\
							&							 & $\hat{P}_f$ & $0.0045$ 	&$0.0020$			 \\
\bottomrule [2pt]
\end{tabular}
\end{table}

\begin{figure}[ht]
\centering
\begin{tabular}{C{7cm}}
\subfloat[]{\includegraphics[width=7cm, height=4.5cm]{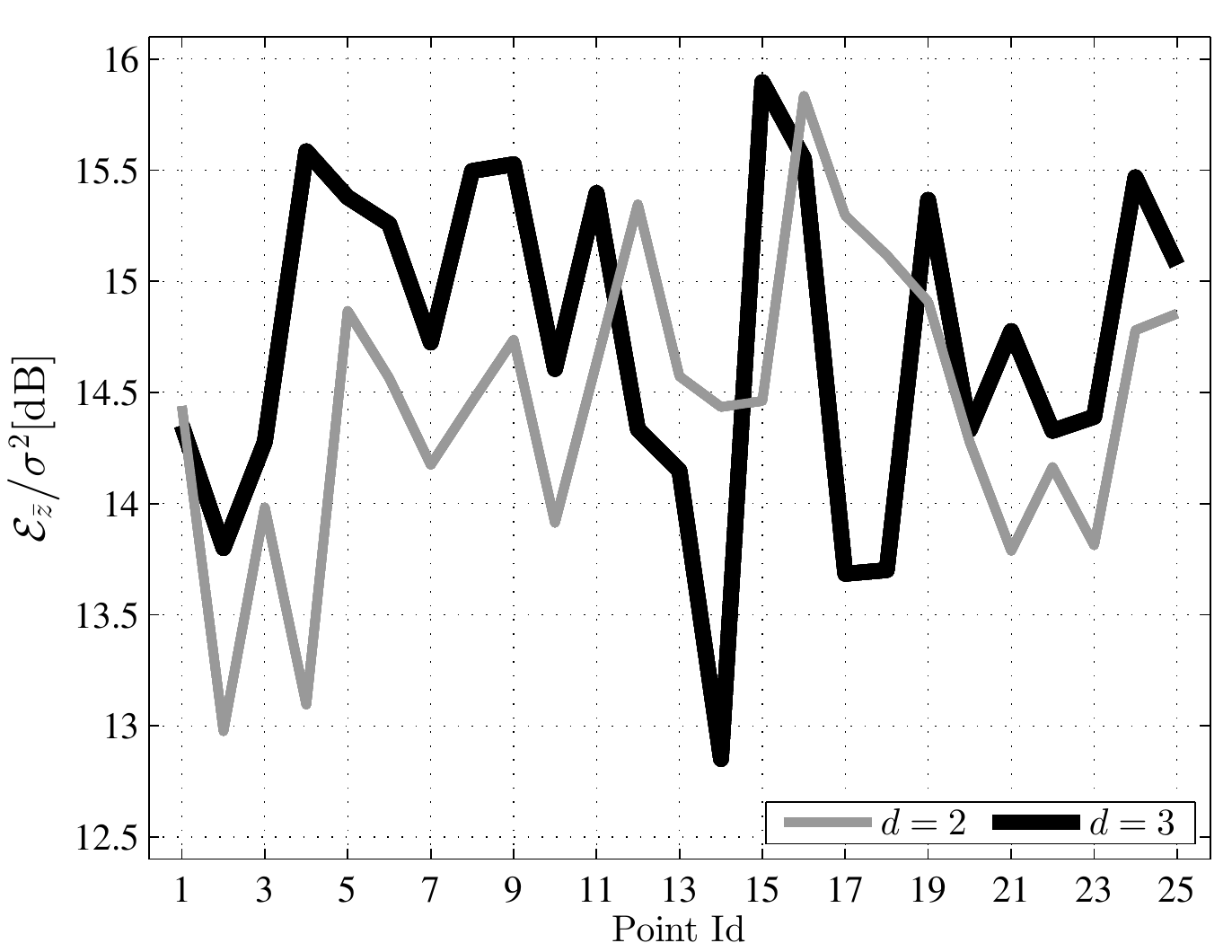}\label{fig:ae_rsa}}\\
\subfloat[]{\includegraphics[width=7cm, height=4.5cm]{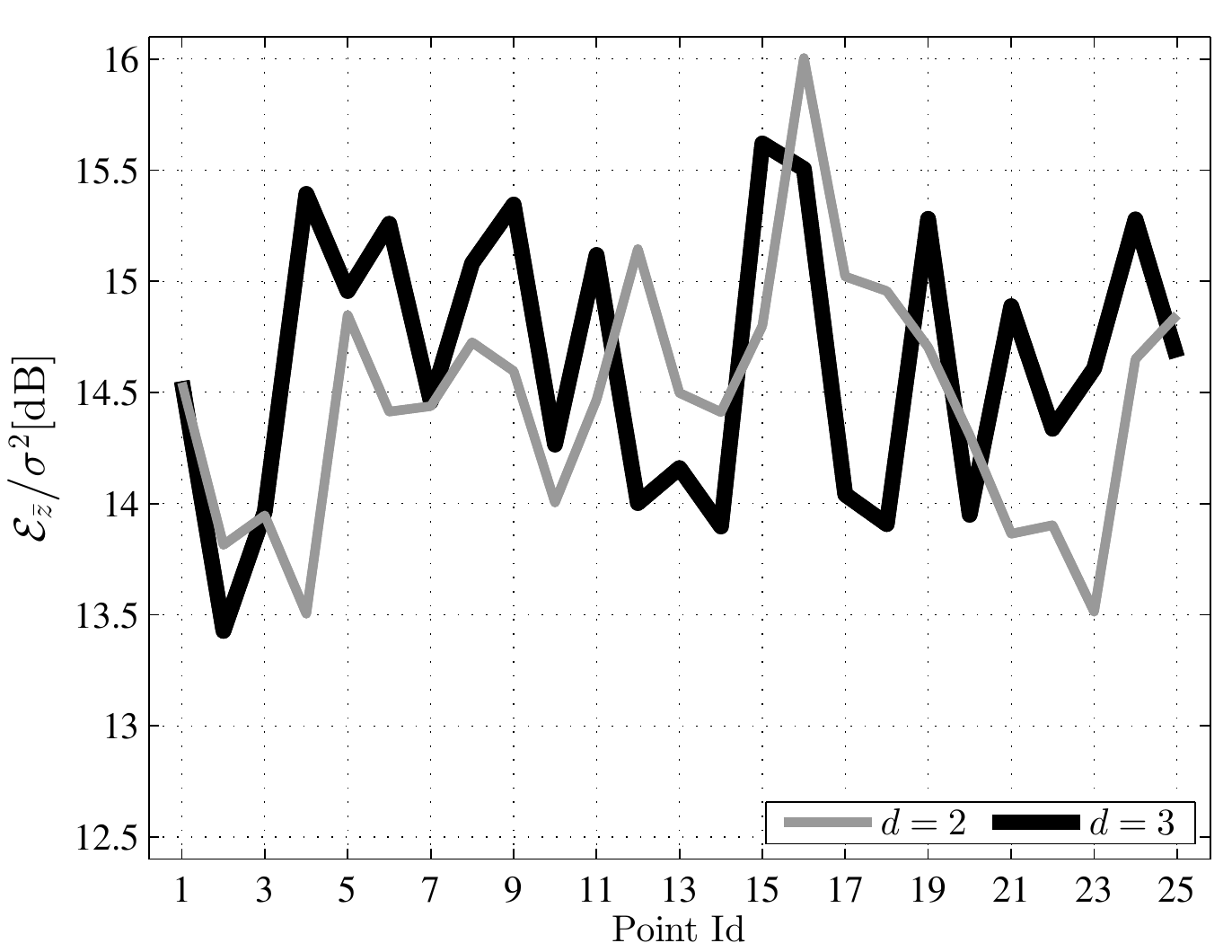}\label{fig:ae_node}} 
\end{tabular}
\caption{Variation of normalized measurement energy with grid point identifier. In (a), using RSA measurement, and in (b) using RX node measurements}
\label{fig:experiment_results1}
\end{figure}

\begin{figure}[ht]
\centering
\begin{tabular}{C{7cm}}
\subfloat[]{\includegraphics[width=7cm, height=4.5cm]{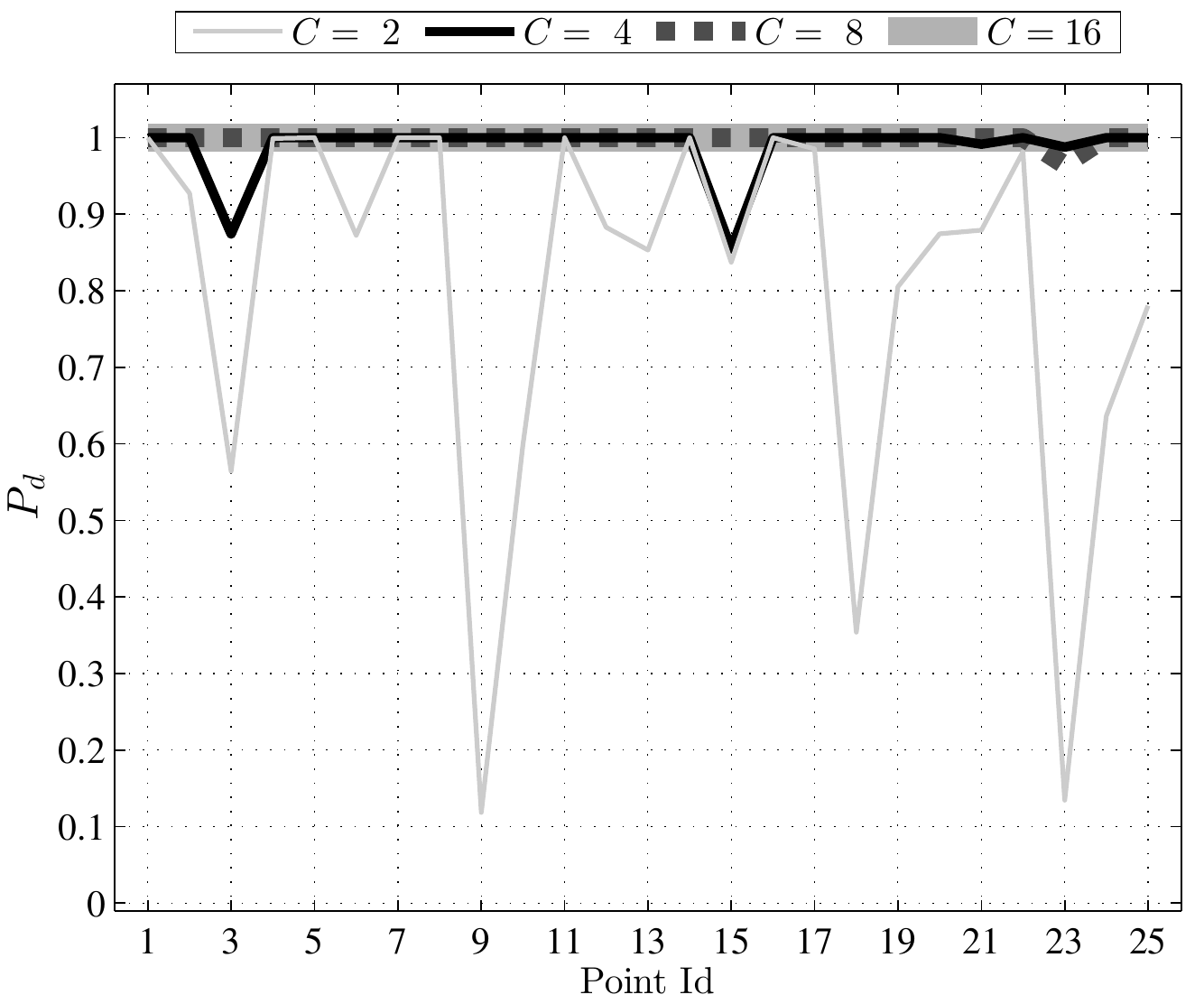}\label{fig:pd_d2}} \\
\subfloat[]{\includegraphics[width=7cm, height=4.5cm]{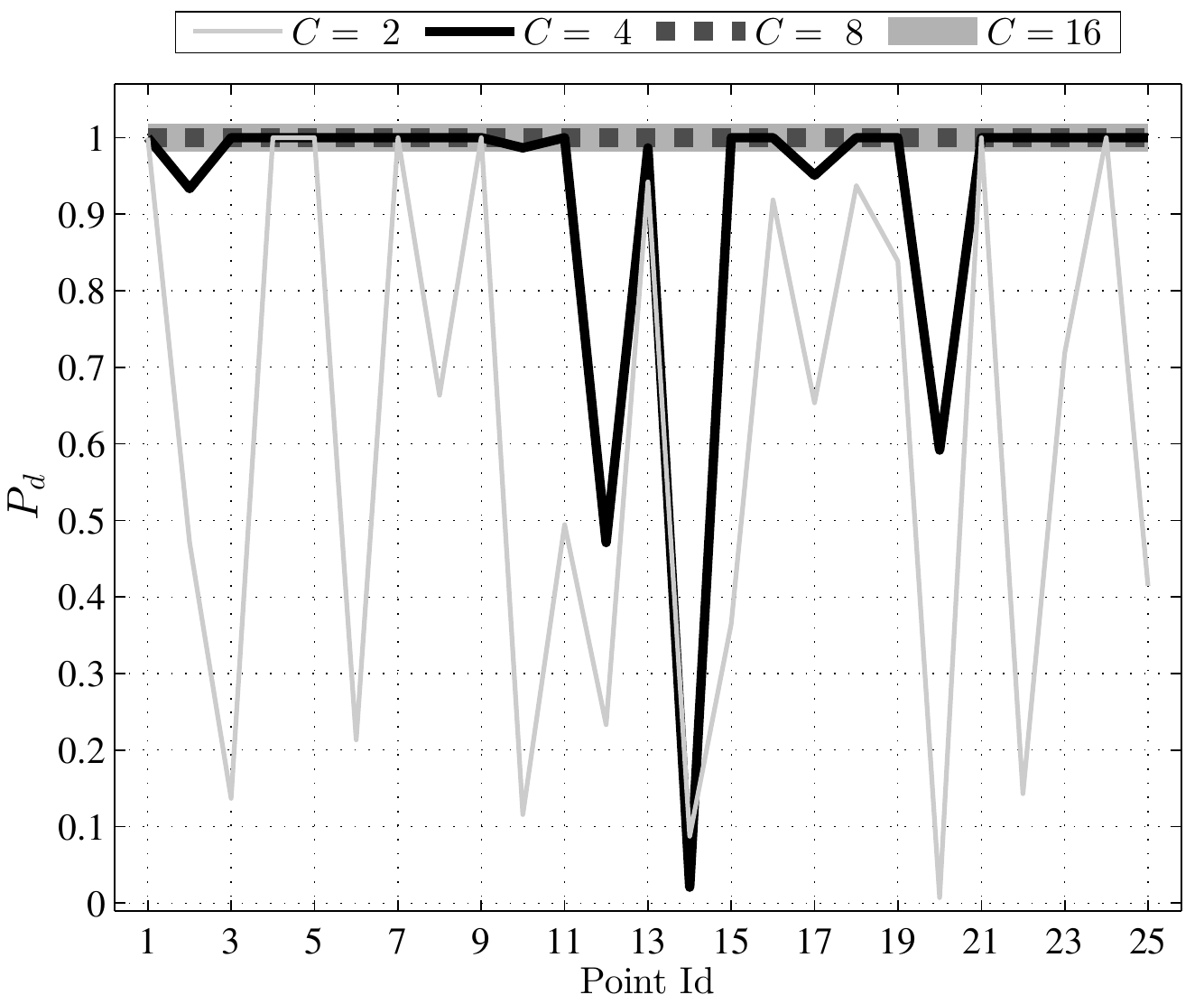}\label{fig:pd_d3}}
\end{tabular}
\caption{Variation of calculated detection probability with number of channels and grid point identifier. In (a),  $d=2 \text{ m}$ experiment, and in (b), $d=3 \text{ m}$ experiment}
\label{fig:experiment_results2}
\end{figure} 

In Table~\ref{table:constant_parameters}, the empirical parameters and false alarm rate results are summarized. The calculated noise variance is denoted by $\sigma$ whereas its increased value is denoted by $\hat{\sigma}$. Using $\hat{\sigma}$ values and $P_f = 6\cdot10^{-6}$, the decision thresholds $X$ are calculated using Eq.~\eqref{eq:threshold} for $C \in \{2, 4, 8, 16\}$. For the given $X$, empty room measurements are used for calculating false alarm rate $\hat{P}_f$. Despite the fact that $\hat{P}_f$ is higher than the specified probability $P_f = 6\cdot10^{-6}$, $\hat{P}_f$ can easily be refined by cross validating the decisions associated with different time samples.  

The normalized average measurement energy  for RSA and RX node measurements as a function of grid point identifier are shown in Fig.~\ref{fig:experiment_results1}. The average measurement is calculated as ensemble mean of all the measurements acquired at each point and using all $16$ channels and is denoted by $\mathcal{E}_{\bar{z}}(\Delta, \Gamma)$. The normalization with estimated noise variance scales the energy values of different experiments to similar ranges. Based on the variance estimates of both RSA and RX node measurements and comparing the results shown in Fig.~\ref{fig:ae_rsa} and Fig.~\ref{fig:ae_node}, it can be concluded that the filtered measurements of RSA and RX node have very similar characteristics. Therefore, we do not give the probability of detection results for RSA measurements due to space limitations.

The probability of detection using the empirical data for different $d$ and $C$ are shown in Fig.~\ref{fig:experiment_results2}. Each graph shows the ratio of correct detections to total number of decisions at each grid point. The channels are selected to keep the mean $f_c$ the same while the frequency separation is selected as large as possible. For example, the energy is calculated using channels $1$ and $16$ for $C=2$, and channels $1,6,11$ and $16$ for $C=4$. The results in Fig.~\ref{fig:experiment_results2} indicate that the larger $d$ is, the better the detection performance is as implied by the definition of $A$ in Eq.~\eqref{eq:a_coefficients}. Regardless of $d$, the detection probability increases as the number of channels increases. Once the number of channels is above $8$, the detection probability is very close to $1$. These results are in coherence with the numerical evaluations in Fig.~\ref{fig:energy_xi_c_d} and Fig.~\ref{fig:pd_c}. To sum up, in order to achieve high probability of detection, the distance between TX and RX should be selected according to size of the monitored area, and the number of utilized frequency channels should be increased as much as possible.  

\section{Discussion}\label{sec:discussion}

The LoS signal synchronization is the most important assumption for validity of the presented derivations starting from Eq.~\eqref{eq:received_power}. The assumption holds when the LoS signal arrives first to the RX antenna and satisfies the conditions to initiate the synchronization sequence. The LoS signal typically fails to initiate the synchronization sequence if it experiences large losses compared to the other multipath components impinging on the RX antenna. This can result if the antenna gain pattern is irregular and the RX is not in the direction of the main lobe or because of the surrounding environment. Failure of the assumption implies that the synchronized multipath component is likely to alter with small changes in the environment and with small variations in the carrier frequency since there are multitudes of non-LoS signals with similar strengths. In this case, the signal strengths in Eq.~\eqref{eq:received_signal_power} are modulated by different realizations of the fading so that associated path-loss exponents cannot be assumed time-invariant and similar. However, in a typical office environment most of the obstructing objects are highly dielectric so that a communication system equipped with high-quality antennas is very likely to synchronize to the LoS signal. Therefore, this assumption is not very restrictive, but requires some attention when selecting the antenna and deploying the system to make sure that the RX synchronizes to LoS signal at all frequency channels. 

\begin{figure}[t]
\centerline{\includegraphics[width=0.85\columnwidth, height=4.2cm]{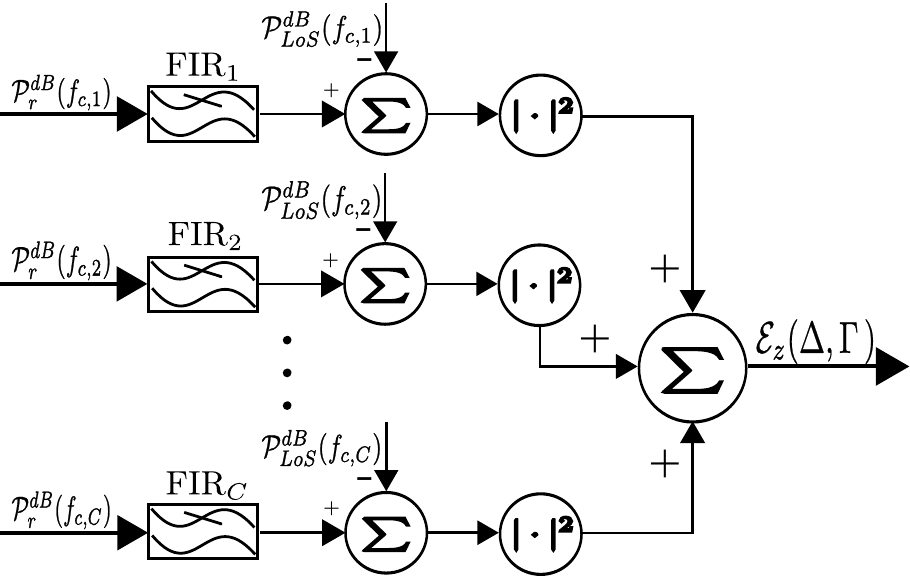}}
\caption{A typical implementation of the detector}
\label{fig:implementation_system}
\end{figure}
A typical implementation of the energy detector requires the components illustrated in Fig.~\ref{fig:implementation_system}. Since the complexity of such a system is very low, it can be embedded in the RX node for local decisions. This leads to reduced communication protocol complexity and enables duty-cycling for low-power operation when the monitored area is vacant. Furthermore, the occupancy output can be refined using the multitudes of decisions associated with different time samples. Therefore, applications requiring occupancy assessment can benefit from very low-cost and low-complexity deployments of narrowband nodes.  

The development efforts of this paper are built upon the assumption that when a person is in the close vicinity of the LoS (without obstructing it), the RSS measurements are altered by human-induced reflections. This assumption is valid for communication systems operating at the $2.4 \text{ GHz}$ ISM band because a person's geometric extent is considerably larger than the wave length. This necessarily implies that single bounce reflections should be accounted for in RSS-based DFL systems operating at the same or higher frequencies resulting to several advantages. First, the measurement models more accurately reflect the RSS changes so that enhanced localization accuracy and resolution are expected. Second, sensitivity region of the wireless links is increased and comparative accuracy with respect to shadowing-based methods can be achieved with fewer nodes. Third, the development of this paper can be utilized in pre-deployment system design and optimization. As an example, the distance between nodes can be determined or optimized \emph{a priori} of deployment to guarantee a demanded spatial coverage with certain detection/false alarm probabilities. This can be achieved by first inverting Eq.~\eqref{eq:roc} and then solving Eq.~\eqref{eq:reflected_energy_approx2} for $d$. Since the ROC is a function of excess path length, for a given false alarm probability, a spatial distribution of detection probability can be calculated. This can for example be used to assign probabilistic pixel weights in imaging based methods.

The effect of human-induced reflections on the RSS model in Eq.~\eqref{eq:effect_of_reflection_db_series} allows one to exchange the parameters such that either frequency or excess path length can be selected as the variable while the other is fixed. In this work, we have considered excess path length as the fixed parameter, and studied the varying frequency. However, if the excess path length is a linear function of time, i.e. a person is moving such that $\Delta = \Delta_0 + \delta t$ where $\delta$ is the time derivative of $\Delta$, for a fixed frequency Eq.~\eqref{eq:effect_of_reflection_db_series} has the same structure. In this case, the energy detector can be used for identifying whether there is a moving person in the environment. However, this approach requires considerably higher SNR which can be achieved using channel diversity. 

Equation~\eqref{eq:effect_of_reflection_db_series} implies that once there is an object approaching to the LoS, the RSS measurements at all frequency channels are altered. This observation can be utilized for discriminating between energy increase in measurements originating from narrowband interference, or presence of a new reflecting object in the environment. Since in both cases the communication system experiences connectivity problems, being able to identify the source allows the network management software to take different countermeasures. In case the connectivity problem is due to interference, it is enough to change the selected communication frequencies in the transmission schedule. On the other hand, if it is due to a new object, the network topology should be altered to resolve the problem. Therefore, the communication frequency parametrization of reflected signal can be used for diagnosing connectivity of wireless links, which is one of the fundamental problems in self-healing wireless networks. 

In this paper, the nodes have direct sequence spread spectrum physical layer. Due to this, the requirement to probe the propagation medium over multiples of frequency channels within the coherence time is a demanding task, and requires a special transmission schedule control in the embedded software. However, communication technologies with frequency hopping and higher number of frequency channels would ease these difficulties by providing already implemented features. As an example, Bluetooth Low-Energy is such a communication standard supporting adaptive frequency hopping over $40$ channels in the same ISM band. The frequency hopping is part of the specification leading to a significant implementation simplification while providing robustness against interference. Therefore, the methods studied in this work are not limited to a specific hardware platform or communication technology, rather there is a broad class of commercial off-the-shelf options.


\section{Conclusions} \label{sec:conclusion}
In this work, detection of a person in a region around a TX-RX pair is studied. The impact of a single bounce human-induced reflection on the RSS measurements is modeled for narrowband coherent receivers. It is shown that the effect is a periodic function of communication frequency, and has closed form Fourier series representation. Based on this finding, an energy detector operating on RSS measurements at different frequency channels is studied. Two approximations of the signal energy are given. The performance of the detector is evaluated by its receiver operating characteristics, and the detection performance is validated with experiments. The studied detector and associated mathematical models have important implications for RSS-based device-free localization systems, including pre-deployment system optimization, sensitivity region assignment for imaging based methods, and aiding classification issues for sequential monte-carlo based tracking/localization methods. Therefore, the available systems allowing multi-channel RSS measurements can benefit from the studied methods to improve their performance or to optimize the deployment.

\appendices
\section{Fourier Series Expansion of Reflected Signal} \label{appendix:fourier_series}

In this appendix, we derive the Fourier series expansion of the reflected signal in logarithmic scale
\begin{equation}\label{eq:effect_of_reflection_db_appendix}
\begin{aligned}
	\zeta(\Delta, \Gamma, \beta) =&~~~ \hat{e} \ln \Big(1 + \Gamma^2 \big({1+\Delta/d}\big)^{-\eta}\Big) \\
			&+\hat{e} \ln \Big(1 -\kappa(\Delta) \cos(\phi) \Big).
\end{aligned}
\end{equation}
It can easily be shown that $\kappa(\Delta) < 1$ for $\Delta \ge 0$ and $\Gamma < 1$, allowing one to write 
\begin{equation*}
	\ln \Big(1 -\kappa(\Delta) \cos(\phi) \Big) = -\sum\limits_{l=1}^{\infty}{\frac{1}{l}\big(\kappa(\Delta) \cos(\phi)\big)^l}. 
\end{equation*}  
Since this is an an even periodic function, its trigonometric Fourier series expansion can be written as 
\begin{equation*}
	\ln \Big(1 -\kappa(\Delta) \cos(\phi) \Big) = b_0 + \sum\limits_{i=1}^{\infty}{b_i \cos(2 \pi i \Delta \beta)},
\end{equation*}
where the coefficients are given by
\begin{equation*}
	\begin{aligned}
	b_i = -\Delta \int\limits_{-\frac{1}{2\Delta}}^{\frac{1}{2\Delta}} {\sum\limits_{l=1}^{\infty}
					{\frac{(\kappa(\Delta))^l}{l}\cos^l(2 \pi \Delta \beta)} \cos(2 \pi i \Delta \beta) d\beta}.
	\end{aligned}
\end{equation*}
In addition, the cosine powers can be expanded as harmonics, 
\begin{equation*}
\begin{aligned}
	\cos^l(\phi) = 
	\begin{cases}
		\frac{2}{2^l}\sum\limits_{i=0}^{\frac{l-1}{2}}{\binom{l}{i}\cos\big( (l-2 i) \phi\big)} & l \text{ odd} \\
		\frac{1}{2^l}\binom{l}{\frac{l}{2}} + \frac{2}{2^l}\sum\limits_{i=0}^{\frac{l-2}{2}}{\binom{l}{i}\cos\big((l-2 i) \phi\big)} & l \text{ even}
	\end{cases}
\end{aligned}
\end{equation*}
Due to orthogonality of the sinusoidal functions, $b_i$ are polynomials of $\kappa(\Delta)$ which can be written as
\begin{equation*}
\begin{aligned}
	b_i = 
	\begin{cases}
		- ~\sum\limits_{l=1}^{\infty}{\frac{\kappa^{2 l}(\Delta)}{2 l 2^{2 l}}}\binom{2 l}{l} & i = 0 \\
		- \sum\limits_{l = \frac{i+1}{2}}^{\infty}{\frac{2 \kappa^{2 l - 1}(\Delta)}{(2 l - 1) 2^{2 l - 1}}}\binom{2 l - 1}{\frac{2 l - i -1}{2}}& i \text{ odd} \\
		- ~\sum\limits_{l = \frac{i}{2}}^{\infty}{\frac{2 \kappa^{2 l }(\Delta)}{(2 l ) 2^{2 l}}}\binom{2 l}{\frac{2 l - i}{2}}& i \text{ even} 
	\end{cases}
\end{aligned}
\end{equation*}
Note that $b_i < 0$ for all $i=0,1,2 \cdots$, and for $\kappa(\Delta)<1$ and $i > 0$ we have $|b_i|>|b_{i+1}|$. The partial sums of  the coefficients are convergent, and after simplification they can be written as 
\begin{equation*} 
	b_i = 
	\begin{cases}
		\ln \left( \frac{1}{2} + \frac{1}{2} \sqrt{1-\kappa^2(\Delta)}\right) & i = 0 \\
		- 2 \frac{\kappa^{i}(\Delta)}{i \left( 1+\sqrt{1-\kappa^2(\Delta)}\right)^i}& i >0 
	\end{cases}
\end{equation*}
Substituting Eq.\eqref{eq:kappa_Delta} into this expression yields
\begin{equation*} \label{eq:fourier_coefficients_appendix}
	b_i = 
	\begin{cases}
		- \ln \left( {1+\Gamma^2 (1+{\Delta}/{d})^{-\eta}}\right) & i = 0 \\
		- \frac{2}{i} \left({\Gamma}{(1+{\Delta}/{d})^{-\frac{\eta}{2}}}\right)^i & i >0 
	\end{cases}
\end{equation*} 
One of the important consequence is the fact that $b_0$ is equal to additive inverse of the first term in Eq.~\eqref{eq:effect_of_reflection_db_appendix}. Therefore, the effect of reflected signal can be written in terms of Fourier series as 
\begin{equation*}\label{eq:effect_of_reflection_db_series_appendix}
	\zeta(\Delta, \Gamma, \beta) = -2 \hat{e} \sum\limits_{i=1}^{\infty}{ \frac{1}{i} \left({\Gamma}{(1+{\Delta}/{d})^{-\frac{\eta}{2}}}\right)^i \cos(2 \pi i \Delta \beta )}.
\end{equation*}

\section{Signal Energy Approximation} \label{appendix:reflected_signal_energy}

The energy of the approximate reflected signal, given in Eq.~\eqref{eq:approximate_signal}, is the square sum of the signals at different channels in $\mathscr{C}$, 
\begin{equation*}
	\mathcal{E}_{\bar{\zeta}}(\Delta, \Gamma) = 4 \hat{e}^2\sum\limits_{l=1}^{C} \Big( a_1 \cos(2\pi \Delta \beta_l) + a_2 \cos(4\pi \Delta \beta_l)\Big)^2,
\end{equation*}
where the squared term can be expanded as
\begin{equation*}
	\begin{aligned}
	\Big( a_1 \cos(2\pi \Delta \beta_l) &+ a_2 \cos(4\pi \Delta \beta_l)\Big)^2 = \frac{a_1^2 + a_2^2}{2} \\
	&  + a_1 a_2 \cos(2 \pi \Delta \beta_l) + \frac{a_1^2}{2} \cos(4 \pi \Delta \beta_l) \\
	& + a_1 a_2 \cos(6 \pi \Delta \beta_l) + \frac{a_2^2}{2} \cos(8 \pi \Delta \beta_l).
	\end{aligned}
\end{equation*}
Let us denote the average inverse wavelength by 
$
\bar{\beta} = \sum_{l=1}^C \beta_l\big/C,
$ 
so that, the sum identity of $\cos(\cdot)$ function,  
\begin{equation*}
\begin{aligned}
	2 \cos(2\pi \Delta \bar{\beta})\cos(2\pi \Delta l \xi) = & \cos\left( 2\pi \Delta(\beta - l \xi) \right) + \\ 
															&\cos\left( 2\pi \Delta(\beta + l \xi) \right),
\end{aligned}
\end{equation*}
can be utilized to find an expression for $\mathcal{E}_{\bar{\zeta}}(\Delta, \Gamma)$. If $C > 0$ is an odd integer, the average channel is in $\mathscr{C}$, and we have 
\begin{equation*}
\begin{aligned}
	\sum_{l=1}^{C} \cos(2 \pi \Delta \beta_l) &= \cos(2\pi \Delta \bar{\beta}) \left(1 + 2 \sum_{l=1}^{\frac{C-1}{2}}\cos(2\pi \Delta l \xi) \right)\\
				&= \cos(2\pi \Delta \bar{\beta})\frac{\sin(\pi \Delta C \xi)}{\sin(\pi \Delta \xi)},
\end{aligned}
\end{equation*}
where the second equality follows from Lagrange trigonometric identity 
\begin{equation*}\label{eq:lagrange_identity}
	\sum_{l=1}^{\frac{C-1}{2}}\cos(2\pi \Delta l \xi) = \frac{1}{2}
			\left( -1 + \frac{\sin(\pi \Delta C \xi)}{\sin(\pi \Delta \xi)}\right).
\end{equation*}  
If $C$ is even, the average channel is not in $\mathscr{C}$, and we have 
\begin{equation*}
\begin{aligned}
	\sum_{l=1}^{C} \cos(2 \pi \Delta \beta_l) = \cos(2\pi \Delta \bar{\beta}) \Bigg(& \cos(\pi \Delta C \xi)-1 + \\
									&\cos(\pi \Delta \xi)\frac{\sin \big(\pi \Delta C \xi \big)}{\sin(\pi \Delta \xi)} \Bigg).
\end{aligned}
\end{equation*}
Using the difference identity of $\sin(\cdot)$ function, we can write
\begin{equation*}
\begin{aligned}
	\sum_{l=1}^{C} \cos(2 \pi \Delta \beta_l) &=& \cos(2\pi \Delta \bar{\beta})&\frac{\sin(\pi \Delta C \xi)}{\sin(\pi \Delta \xi)} \Bigg( \sin(\pi \Delta \xi) + \\
									&&&\tan \left(\pi \Delta \frac{C}{2} \xi \right)  \cos(\pi \Delta \xi)\Bigg) \\
									&\approx& \cos(2\pi \Delta \bar{\beta})&\frac{\sin(\pi \Delta C \xi)}{\sin(\pi \Delta \xi)},
\end{aligned}
\end{equation*}
where the approximation is valid for $\Delta \le 1\big/ (2 C \xi)$. Since for $\Delta > 1\big/ (2 C \xi)$ the amplitude of these terms is small, this approximation can be utilized for all $\Delta \ge 0$. Therefore, for odd $C$, $\mathcal{E}_{\bar{\zeta}}(\Delta, \Gamma)$ is given in Eq.~\eqref{eq:reflected_energy}, and for even $C$ the result is similar, and can be approximated with Eq.~\eqref{eq:reflected_energy}.

\section{Characteristic Function of Gaussian Quadratic Form}\label{appendix:quadratic_characteristic_equation}
In this appendix, we give the characteristic function of general real valued quadratic function in the form
\begin{equation*}
	{y} = \boldsymbol{x}^T \boldsymbol{Q} \boldsymbol{x},
\end{equation*}
where $\boldsymbol{Q}$ is a positive definite matrix in $\mathbb{R}^{N \times N}$, and $\boldsymbol{x}$ is a multivariate Gaussian in $\mathbb{R}^N$ of components $x_i$ with mean $\bar{\boldsymbol{x}}$ and positive definite covariance $\boldsymbol{\Sigma}$. The characteristic function of $y$ is given by
\begin{equation}\label{eq:quadratic_form_characteristic_function}
	\psi(\omega) = \frac{
				 \exp\left({-\frac{1}{2}\bar{\boldsymbol{x}}^T \boldsymbol{\Sigma}^{-1}
					\left( \boldsymbol{I} - \left( \boldsymbol{I} - j 2 \omega \boldsymbol{Q} \boldsymbol{\Sigma}\right)^{-1}\right)\bar{\boldsymbol{x}}}\right)}
					{|\boldsymbol{I}- j 2 \omega \boldsymbol{Q} \boldsymbol{\Sigma} |^{{2}}},
\end{equation}
where $|\cdot|$ denotes matrix determinant, and $\boldsymbol{I}$ is the identity matrix in $\mathbb{R}^{N \times N}$. This expression can be deduced from the more general form for complex multivariate Gaussian case derived by Turin \cite{Turin1960}.

For the purposes of this paper, we are interested in the distribution of $y$ for $\boldsymbol{Q} = \boldsymbol{I}$ and diagonal covariance matrix $\boldsymbol{\Sigma} = \diag\{ \sigma_1^2, \cdots, \sigma_N^2\}$ ($x_i$ are uncorrelated). First, suppose $\bar{\boldsymbol{x}} \ne \boldsymbol{0}$, then we have 
\begin{equation}\label{eq:psi_omega}
	\psi(\omega) = \prod\limits_{i=1}^N \frac{\exp\left({ \frac{j \omega \bar{x}_i^2}{1-j 2 \omega \sigma_i^2}}\right)}{\left(1 - j 2 \omega \sigma_i^2\right)^\frac{1}{2}},
\end{equation} 
where $\bar{x}_i$ are the components of $\bar{\boldsymbol{x}}$. If $x_i$ have identical variances, i.e. $\boldsymbol{\Sigma} = \sigma^2\boldsymbol{I}$, we have characteristic function of non-central chi-square distribution with non-centrality parameter $\mu = \sum \bar{x}_i^2$ of $N$ degrees of freedom. Second, suppose that $\bar{\boldsymbol{x}} = \boldsymbol{0}$, then Eq.~\eqref{eq:psi_omega} simplifies to
the characteristic function of sum of $N$ independent Gamma variates with equal shape parameter $1/2$ and different scale parameters ($2\sigma_i^2$). Infinite series representation of the distribution of such a variate has been derived by Moschopoulos \cite{Moschopoulos1985}. If we also suppose that   $\boldsymbol{\Sigma} = \sigma^2\boldsymbol{I}$, then we reach the characteristic function of a Gamma variate with shape parameter $N/2$ and scale parameter ($2\sigma^2$), which is equivalent to central chi-square distribution of $N$ degrees of freedom. 


\ifCLASSOPTIONcaptionsoff
  \newpage
\fi

\bibliographystyle{IEEEtran}

\begin{thebibliography}{10}
\providecommand{\url}[1]{#1}
\csname url@samestyle\endcsname
\providecommand{\newblock}{\relax}
\providecommand{\bibinfo}[2]{#2}
\providecommand{\BIBentrySTDinterwordspacing}{\spaceskip=0pt\relax}
\providecommand{\BIBentryALTinterwordstretchfactor}{4}
\providecommand{\BIBentryALTinterwordspacing}{\spaceskip=\fontdimen2\font plus
\BIBentryALTinterwordstretchfactor\fontdimen3\font minus
  \fontdimen4\font\relax}
\providecommand{\BIBforeignlanguage}[2]{{%
\expandafter\ifx\csname l@#1\endcsname\relax
\typeout{** WARNING: IEEEtran.bst: No hyphenation pattern has been}%
\typeout{** loaded for the language `#1'. Using the pattern for}%
\typeout{** the default language instead.}%
\else
\language=\csname l@#1\endcsname
\fi
#2}}
\providecommand{\BIBdecl}{\relax}
\BIBdecl

\bibitem{Patwari2010}
N.~Patwari and J.~Wilson, ``Rf sensor networks for device-free localization:
  Measurements, models, and algorithms,'' \emph{Proceedings of the IEEE},
  vol.~98, no.~11, pp. 1961--1973, 2010.

\bibitem{Kaltiokallio2014}
O.~Kaltiokallio, H.~Yigitler, R.~J{\"a}ntti, and N.~Patwari, ``Non-invasive
  respiration rate monitoring using a single {COTS} {TX}-{RX} pair,'' in
  \emph{Information Processing in Sensor Networks, 2014. IPSN '14.
  International Conference on}, 2014, to appear.

\bibitem{Wilson2010}
J.~Wilson and N.~Patwari, ``Radio tomographic imaging with wireless networks,''
  \emph{IEEE Transactions on Mobile Computing}, vol.~9, no.~5, pp. 621--632,
  May 2010.

\bibitem{Kaltiokallio2012a}
O.~Kaltiokallio, M.~Bocca, and N.~Patwari, ``Follow@ grandma: Long-term
  device-free localization for residential monitoring,'' in \emph{37th Annual
  IEEE Conference on Local Computer Networks -- Workshops}.\hskip 1em plus
  0.5em minus 0.4em\relax IEEE, Oct. 2012, pp. 991--998.

\bibitem{Saeed2014}
A.~Saeed, A.~Kosba, and M.~Youssef, ``Ichnaea: A low-overhead robust wlan
  device-free passive localization system,'' \emph{Selected Topics in Signal
  Processing, IEEE Journal of}, vol.~8, no.~1, pp. 5--15, Feb 2014.

\bibitem{Geng2014}
L.~Geng, M.~Bugallo, A.~Athalye, and P.~Djuric, ``Indoor tracking with rfid
  systems,'' \emph{Selected Topics in Signal Processing, IEEE Journal of},
  vol.~8, no.~1, pp. 96--105, Feb 2014.

\bibitem{Bocca2013}
M.~Bocca, O.~Kaltiokallio, and N.~Patwari, ``Radio tomographic imaging for
  ambient assisted living,'' in \emph{Evaluating AAL Systems Through
  Competitive Benchmarking}.\hskip 1em plus 0.5em minus 0.4em\relax Springer,
  2013, pp. 108--130.

\bibitem{Bjorkbom2013}
M.~Bjorkbom, J.~Timonen, H.~Yigitler, O.~Kaltiokallio, J.~Garcia, M.~Myrsky,
  J.~Saarinen, M.~Korkalainen, C.~Cuhac, R.~Jantti, R.~Virrankoski, J.~Vankka,
  and H.~Koivo, ``Localization services for online common operational picture
  and situation awareness,'' \emph{Access, IEEE}, vol.~1, pp. 742--757, 2013.

\bibitem{Anderson2014}
C.~Anderson, R.~Martin, T.~Walker, and R.~Thomas, ``Radio tomography for
  roadside surveillance,'' \emph{Selected Topics in Signal Processing, IEEE
  Journal of}, vol.~8, no.~1, pp. 66--79, Feb 2014.

\bibitem{Li2011}
Y.~Li, X.~Chen, M.~Coates, and B.~Yang, ``Sequential monte carlo
  radio-frequency tomographic tracking,'' in \emph{Acoustics, Speech and Signal
  Processing (ICASSP), 2011 IEEE International Conference on}, 2011, pp.
  3976--3979.

\bibitem{Vaughan2003}
R.~Vaughan, \emph{Channels, propagation and antennas for mobile
  communications}.\hskip 1em plus 0.5em minus 0.4em\relax Iet, 2003, no.~50.

\bibitem{Agrawal2009}
P.~Agrawal and N.~Patwari, ``Correlated link shadow fading in multi-hop
  wireless networks,'' \emph{Wireless Communications, IEEE Transactions on},
  vol.~8, no.~8, pp. 4024--4036, 2009.

\bibitem{Kaltiokallio2014a}
O.~Kaltiokallio, H.~Yi{\u{g}}itler, and R.~J{\"a}ntti, ``A three-state received
  signal strength model for device-free localization,'' \emph{arXiv preprint
  arXiv:1402.7019}, 2014.

\bibitem{Savazzi2014}
S.~Savazzi, M.~Nicoli, F.~Carminati, and M.~Riva, ``A bayesian approach to
  device-free localization: Modeling and experimental assessment,''
  \emph{Selected Topics in Signal Processing, IEEE Journal of}, vol.~8, no.~1,
  pp. 16--29, Feb 2014.

\bibitem{Patwari2011}
N.~Patwari and J.~Wilson, ``Spatial models for human motion-induced signal
  strength variance on static links,'' \emph{Information Forensics and
  Security, IEEE Transactions on}, vol.~6, no.~3, pp. 791--802, 2011.

\bibitem{LinkLineVideo}
\BIBentryALTinterwordspacing
{A three-state RSS model for DFL}. [Online]. Available:
  \url{https://sites.google.com/site/ossikaltiokallio/research/rss-based-dfl}
\BIBentrySTDinterwordspacing

\bibitem{Molisch2011}
A.~F. Molisch, \emph{Wireless communications}, 2nd~ed.\hskip 1em plus 0.5em
  minus 0.4em\relax John Wiley \& Sons, 2011, vol.~15.

\bibitem{Naghiyev2014}
E.~Naghiyev, M.~Gillott, and R.~Wilson, ``Three unobtrusive domestic occupancy
  measurement technologies under qualitative review,'' \emph{Energy and
  Buildings}, vol.~69, pp. 507--514, 2014.

\bibitem{Lin2011}
W.-C. Lin, W.-G. Seah, and W.~Li, ``Exploiting radio irregularity in the
  internet of things for automated people counting,'' in \emph{Personal Indoor
  and Mobile Radio Communications (PIMRC), 2011 IEEE 22nd International
  Symposium on}, Sept 2011, pp. 1015--1019.

\bibitem{Mrazovac2012}
B.~Mrazovac, M.~Bjelica, D.~Kukolj, B.~Todorovic, and N.~Teslic, ``A human
  detection method for residential smart energy systems based on zigbee rssi
  changes,'' in \emph{Consumer Electronics (ICCE), 2012 IEEE International
  Conference on}, Jan 2012, pp. 110--111.

\bibitem{Yigitler2013}
H.~Yigitler, O.~Kaltiokallio, and R.~Jantti, ``A management framework for
  device-free localization,'' in \emph{Neural Networks (IJCNN), The 2013
  International Joint Conference on}, Aug 2013, pp. 1--8.

\bibitem{Kaltiokallio2012}
O.~Kaltiokallio, M.~Bocca, and N.~Patwari, ``Enhancing the accuracy of radio
  tomographic imaging using channel diversity,'' in \emph{the 9th IEEE
  International Conference on Mobile Ad hoc and Sensor Systems}, 2012.

\bibitem{Skolnik2008}
M.~I. Skolnik, ``An introduction and overview of radar,'' in \emph{Radar
  Handbook}, M.~I. Skolnik, Ed.\hskip 1em plus 0.5em minus 0.4em\relax
  McGraw-Hill, 2008, ch.~1.

\bibitem{Paolini2008}
E.~Paolini, A.~Giorgetti, M.~Chiani, R.~Minutolo, and M.~Montanari,
  ``Localization capability of cooperative anti-intruder radar systems,''
  \emph{EURASIP Journal on Advances in Signal Processing}, vol. 2008, no.~17,
  2008.

\bibitem{Kilic2014}
Y.~Kilic, H.~Wymeersch, A.~Meijerink, M.~Bentum, and W.~Scanlon, ``Device-free
  person detection and ranging in uwb networks,'' \emph{Selected Topics in
  Signal Processing, IEEE Journal of}, vol.~8, no.~1, pp. 43--54, Feb 2014.

\bibitem{Zhou2013}
Z.~Zhou, Z.~Yang, C.~Wu, L.~Shangguan, and Y.~Liu, ``Towards omnidirectional
  passive human detection,'' in \emph{INFOCOM, 2013 Proceedings IEEE}.\hskip
  1em plus 0.5em minus 0.4em\relax IEEE, 2013, pp. 3057--3065.

\bibitem{Kosba2012}
A.~E. Kosba, A.~Saeed, and M.~Youssef, ``Rasid: A robust wlan device-free
  passive motion detection system,'' in \emph{Pervasive Computing and
  Communications (PerCom), 2012 IEEE International Conference on}.\hskip 1em
  plus 0.5em minus 0.4em\relax IEEE, 2012, pp. 180--189.

\bibitem{Mrazovac2013}
B.~Mrazovac, B.~Todorovi{\'c}, M.~Bjelica, and D.~Kukolj, ``Device-free indoor
  human presence detection method based on the information entropy of rssi
  variations,'' \emph{Electronics Letters}, vol.~49, no.~22, pp. 1386--1388,
  2013.

\bibitem{Wilson2012}
J.~Wilson and N.~Patwari, ``A fade-level skew-laplace signal strength model for
  device-free localization with wireless networks,'' \emph{IEEE Transactions on
  Mobile Computing}, vol.~11, no.~6, pp. 947--958, Jun. 2012.

\bibitem{Zheng2012}
Y.~Zheng and A.~Men, ``Through-wall tracking with radio tomography networks
  using foreground detection,'' in \emph{Wireless Communications and Networking
  Conference (WCNC), 2012 IEEE}.\hskip 1em plus 0.5em minus 0.4em\relax IEEE,
  2012, pp. 3278--3283.

\bibitem{Martin2014}
R.~Martin, A.~Folkerts, and T.~Heinl, ``Accuracy vs. resolution in radio
  tomography,'' \emph{Signal Processing, IEEE Transactions on}, vol.~PP,
  no.~99, pp. 1--1, 2014, to appear.

\bibitem{Pinel2010}
N.~Pinel, C.~Bourlier, and J.~Saillard, ``Degree of roughness of rough layers:
  Extensions of the rayleigh roughness criterion and some applications,''
  \emph{Progress In Electromagnetics Research B}, vol.~19, pp. 41--63, 2010.

\bibitem{Rappaport2002}
T.~S. Rappaport, \emph{Wireless Communications: Principles and Practice},
  2nd~ed.\hskip 1em plus 0.5em minus 0.4em\relax Prentice Hall, 2002.

\bibitem{Sankaralingam2010}
S.~Sankaralingam and B.~Gupta, ``Determination of dielectric constant of fabric
  materials and their use as substrates for design and development of antennas
  for wearable applications,'' \emph{Instrumentation and Measurement, IEEE
  Transactions on}, vol.~59, no.~12, pp. 3122--3130, 2010.

\bibitem{TissueElectric}
``Calculation of the dielectric properties of body tissues in the frequency
  range 10 hz - 100 ghz,'' \url{http://niremf.ifac.cnr.it/tissprop/}, 2013.

\bibitem{Proakis2008}
J.~G. Proakis and M.~Salehi, \emph{Digital Communications}, 5th~ed.\hskip 1em
  plus 0.5em minus 0.4em\relax New York, NY: McGraw-Hill, 2008.

\bibitem{Maximon2003}
L.~C. Maximon, ``The dilogarithm function for complex argument,''
  \emph{Proceedings of the Royal Society of London. Series A: Mathematical,
  Physical and Engineering Sciences}, vol. 459, no. 2039, pp. 2807--2819, 2003.

\bibitem{Trees1968}
H.~V. Trees, \emph{Detection, estimation, and modulation theory, Part I}.\hskip
  1em plus 0.5em minus 0.4em\relax Jhon Wiley and Sons, 1968.

\bibitem{Urkowitz1967}
H.~Urkowitz, ``Energy detection of unknown deterministic signals,''
  \emph{Proceedings of the IEEE}, vol.~55, no.~4, pp. 523--531, 1967.

\bibitem{Kostylev2000}
V.~Kostylev, ``Characteristics of energy detection of quasideterministic radio
  signals,'' \emph{Radiophysics and Quantum Electronics}, vol.~43, no.~10, pp.
  833--839, 2000.

\bibitem{Digham2007}
F.~F. Digham, M.-S. Alouini, and M.~K. Simon, ``On the energy detection of
  unknown signals over fading channels,'' \emph{IEEE Transactions on
  Communications}, vol.~55, no.~1, pp. 21--24, 2007.

\bibitem{Abramowitz1970}
M.~Abramowitz and I.~A. Stegun, \emph{Handbook of mathematical functions: with
  formulas, graphs, and mathematical tables}.\hskip 1em plus 0.5em minus
  0.4em\relax Dover Publications, 1970.

\bibitem{Turin1960}
G.~L. Turin, ``The characteristic function of hermitian quadratic forms in
  complex normal variables,'' \emph{Biometrika}, vol.~47, no. 1-2, pp.
  199--201, 1960.

\bibitem{Moschopoulos1985}
P.~Moschopoulos, ``The distribution of the sum of independent gamma random
  variables,'' \emph{Annals of the Institute of Statistical Mathematics},
  vol.~37, no.~1, pp. 541--544, 1985.

\end{thebibliography}

\end{document}